\documentclass[preprint2]{aastex6}
\usepackage{epsfig}
\usepackage{amsfonts,amssymb,amsmath}

\begin{document}

\title{The Roles of Mass and Environment in the Quenching of Galaxies}
\author{E. Contini$^{1}$, Q. Gu$^{1}$, X. Kang$^{2}$, J. Rhee$^{3}$, S.K. Yi$^{3}$}

\affil{$^1$School of Astronomy and Space Science, Nanjing University, Nanjing 210093, China}
\affil{$^2$Purple Mountain Observatory, the Partner Group of MPI f\"ur Astronomie, 2 West Beijing Road, Nanjing 210008, China}
\affil{$^3$Department of Astronomy and Yonsei University Observatory, Yonsei University, Yonsei-ro 50, Seoul 03722, Republic of Korea}

\email{emanuele.contini82@gmail.com}

\begin{abstract} 
We study the roles of stellar mass and environment in quenching the star formation activity of a large set of simulated galaxies 
by taking advantage of an analytic model coupled to the merger tree extracted from an N-body simulation. The analytic 
model has been set to match the evolution of the global stellar mass function since redshift $z\sim 2.3$ and give reasonable predictions 
of the star formation history of galaxies at the same time. 
We find that stellar mass and environment play different roles: the star formation rate/specific star formation rate-$M_*$ relations are 
independent of the environment (defined as the halo mass) at any redshift probed, $0<z<1.5$, for both star forming and quiescent galaxies, while 
the star formation rate-$M_{halo}$ relation strongly depends on stellar mass in the same redshift range, for both star forming and quiescent galaxies. 
Moreover, the star formation rate and the specific star formation rate are strongly dependent on stellar mass even when the distance from the 
cluster core is used as a proxy for the environment, rather than the halo mass. We then conclude that stellar mass is the main driver of galaxy 
quenching at any redshift probed in this study, not just at $z>1$ as generally claimed, while the environment has a minimal role. All the physical 
processes linked to the environment must act on very short timescales, such that they do not influence the star formation of active galaxies, but 
increase the probability of a given galaxy to become quiescent.

\end{abstract}


\section[]{Introduction} 
\label{sec:intro}
Galaxies are an important component of the visible matter in the Universe. Given the diversity of their morphologies and general properties, they evolve as a consequence of several 
physical processes which are responsible for the different populations that we can observe in the local Universe. A deep understanding of these processes, in particular the role of quenching
and the time/mass-scales involved, would end up in a significant step forward in the comprehension of galaxy formation and evolution. 

It is well known that, broadly speaking, galaxies can be classified into two main populations according to their rate of star formation activity: star forming systems and quiescent (passive) objects
(\citealt{blanton03,baldry04,kauffmann04,brinchmann04,balogh04,cassata08,pallero18,davies19}). Star forming galaxies actively form new stars, have blue colors, late-type morphologies and are 
typically young (\citealt{blanton03,kauffmann03,noeske07,wuyts11}). On the other hand, quiescent galaxies do not show star formation activity, have red colors, early-type morphologies and are 
typically old (\citealt{baldry04,gallazzi08,wetzel12,vanderwel14}).

Galaxy properties are also found to be both environment and stellar mass dependent. Generally speaking, galaxies in denser environment tipically have early-type morphologies, are less star forming, 
redder, older and more metal rich (\citealt{dressler80,kauffmann04,vonderlinden10,peng10,cooper10}), and the same trends are still valid for more massive galaxies (\citealt{kauffmann03,baldry06,weinmann06,bamford09,peng10}).
Environment and stellar mass have been found to be important for the quenching of galaxies, although we do not have a clear knowledge yet of which between environment and mass plays the most important role 
in galaxy quenching (sometimes it is referred to as \emph{nature/nurture} debate).

During the past years, many physical processes related to both environment and stellar mass have been invoked in order to explain galaxy quenching (\citealt{noeske07,peng10,sobral11,muzzin12,muzzin13,darvish16,trussler18}).
In their pioneering work \cite{peng10}, who used SDSS and zCOSMOS data, demonstrated the mutual independence of stellar mass and environment in quenching star formation. From the empirical model they constructed,
they have been able to separate the effects of mass and environmental quenching, and found that mass quenching is the main process responsible for quenching star formation in galaxies with   
$\log M_* > 10.6$, independently of environment and redshift. On the other hand, environmental processes become important at low redshift and for low-mass galaxies. In short, massive galaxies are more likely 
quenched by internal processes that are independent of the environment in which they reside, and galaxies in denser environment are likely quenched by processes that are independent of their stellar mass.

Mass quenching is generelly referred to internal processes that mainly depend on the galaxy mass. Different processes have been proposed depending on the characteristic stellar mass regime. In the low mass 
regime ($\log M_* <9$) gas outflows driven by stellar feedback such as stellar winds/radiation or SNe explosions are thought to play an important part in quenching star formation (\citealt{larson74,dekelsilk86,dallavecchiaschaye08}).
For more massive galaxies ($\log M_* >10$), in particular those with a pronunced bulge component, AGN feedback appears to be more effective in stopping star formation. The AGN can be powerful enough to either 
heat up the surrounding cold gas by injecting energy via radio jets or winds, or even sweep away the gas content through powerful outflows (\citealt{croton06,fabian12,fang13,cicone14,bremer18}).

Environmental quenching is usually intended as the process, or series of processes, that quench star formation because of interactions between galaxies and their surroundings, such as ram pressure stripping 
(\citealt{gunngott72,poggianti17}), strangulation or starvation (\citealt{larson80,moore99}) and harassments (\citealt{faroukishapiro81,moore96}). Ram pressure stripping in clusters removes the cold gas in the 
interstellar medium (ISM) due to the interaction between it and the intracluster medium, thus inhibiting further star formation unless hot gas can cool and replenish the cold gas reservoir. Starvation (or strangulation)
is a process which is assumed instantantaneous as soon as a galaxy is accreted in a large system and that completely removes the hot gas available for cooling, thus shutting down the fuel for further star 
formation. Harassments are instead the result of close galaxy-galaxy encounters which can lead to the removal of gas and the conversion of part of the cold gas into stars.  

All the above mentioned mass/environmental processes can be otherwise classified as processes that act on \emph{central} galaxies (mass quenching), and on \emph{satellite} galaxies (environmental quenching). 
Centrals are either field galaxies or the most massive galaxies residing in the centre of groups/clusters, while satellites were formerly centrals and became satellites once accreted in larger system. This central/satellite 
dichotomy has often been used (especially by the theoretical side) as a parallelism with mass/environmental quenching (e.g. \citealt{vandenbosch08,peng12,wetzel13,contini17b}).

In order to understand what quenching dominates during the evolutionary history of galaxies, it is necessary to separate their contributions. In the past few years, many studies focused on this point (e.g., 
\citealt{kauffmann03,muzzin12,koyama13,darvish16,lagana18} and references therein), in understanding how the star formation rate (SFR) or colors depend on halo mass/clustercentric distance at fixed stellar mass, which 
quantifies mass quenching, and how the SFR-$M_*$ relations vary as a function of environment, which quantifies environmental quenching, at different redshifts. Although we know that stellar mass does play a role, the 
picture is not yet clear for what concerns the environment. A bunch of studies have found that galaxies are more likely to be quenched or red in more massive haloes (see e.g., \citealt{balogh00,depropris04,weinmann06,blantonberlind07,kimm09}), but others 
(e.g., \citealt{pasquali09,vulcani10,muzzin12,koyama13,darvish16,lagana18}) have found no or little dependence on either halo mass or clustercentric distance.

In this paper we make use of the analytic model described in \cite{contini17a,contini17b} coupled with a merger tree constructed from a high-resolution N-body simulation. The model has been developed in order 
to match the stellar mass function at high redshift and predict its evolution with time, with an average ($1-\sigma$) precision  $<0.1$ dex in over three orders of magnitudes in stellar mass at $z\sim 0.3$. 
Our model treats the quenching of star formation according to an exponential decay of the star formation rate with time, which depends on several galaxy properties such as stellar mass or type (satellite/central).
Environment and mass quenching are hence already implemented in our model. The primary goal of this paper is to identify the main quenching mode (mass or environment) as a function of redshift, and 
compare our results with those available in the literature.

The manuscript is structured as follow. In Section \ref{sec:methods} we describe the main features of our model and simulation. In Section \ref{sec:results} we present our results which will be fully discussed 
in Section \ref{sec:discussion}. In Section \ref{sec:conclusions} we summarize the main conclusions of our analysis. Throughout this paper we use a standard cosmology, namely: 
$\Omega_{\lambda}=0.73, \Omega_{m}=0.27, \Omega_{b}=0.044$, $h=0.7$ and $\sigma_{8}=0.81$. Stellar masses are computed by assuming a \cite{chabrier03} Initial Mass Function (IMF).

\section[]{Methods}  
\label{sec:methods}
In the following analysis we use the prediction of an analytic model developed in \cite{contini17a} and refined in \cite{contini17b}. We refer the readers to those papers for the details of the physics implemented
and here we briefly describe the main features. The model has run on the merger tree of an N-Body simulation, whose characteristics are fully mentioned in \cite{kang12}, and shortly summarized in \cite{contini17a}.

The analytic model uses the so-called subhalo abundance matching (ShAM) technique to populate dark matter haloes with galaxies (e.g. \citealt{vale04}), and its main goal is to predict the evolution of the global 
galaxy stellar mass function (SMF). For this purpose, the model is forced to match the observed SMF at $z_{match}=2.3$, such that the predicted and observed SMF are the same. By reading the merger tree of the N-Body 
simulation, the model sorts dark matter haloes and at each time assigns a galaxy to each halo according to the stellar mass-halo mass relation valid at that particular redshift. Once galaxies are set, they evolve 
according to their merger histories which are given by the merger tree, and to their star formation histories. The evolution of the SFR is the novelty in our model. At $z_{match}$ (or the redshift 
when they are born if $z_{form} < z_{match}$) a SFR is assigned to each galaxy by means of the SFR-$M_*$ relation observed at that redshift, and the SFR will evolve down to the present time (unless the galaxy merges
or is dispruted) according to the $\tau$ model described in \cite{contini17b}, depending on the galaxy type (central or satellite).  Moreover, due to gravitational interactions with their host halos, 
galaxies might lose a given amount of stellar mass once they are accreted in larger system (i.e. they become satellites). The model does consider stellar stripping \footnote{Stellar stripping has been proved to be 
the main channel for the formation of the intracluster light in galaxy groups and clusters. For further detail on this topic, see \cite{contini14,contini18,contini19} and references therein.} and the details of the 
implementation can be found in \cite{contini17a}.

\subsection[]{Mass and Environmental Quenching Prescriptions} 
\label{sec:quenching}
For the purposes of this paper, it is worth to fully describe the decay with time of the SFR ($\tau$ model), for both central and satellite galaxies, since it basically accounts for the mass and environmental quenching.
As explained above, the model first assigns an SFR to each galaxy according to the SFR$-M_*$ relation either at $z=z_{match}$ or at $ z=z_{form}$, in case a galaxy forms after $ z_{match}$. From that redshift on, the
SFR evolves according to functional forms that consider information such as type (central or satellite), stellar mass, and a quenching timescale. The star formation histories (SFHs) of centrals and satellites are treated 
separately. For centrals, we use a prescription very similar to the one adopted in \cite{noeske07}:

\begin{equation} \label{eq:sfrcen}
SFR_{cen}(t)=SFR_{match/form} \cdot \exp{\left(-\frac{t}{\tau_c}\right)} \, ,
\end{equation}
where $\tau_c$ is the quenching timescale of centrals. $\tau_c$ is derived from the following equation:

\begin{equation} \label{eq:taus}
\tau_{c} = 10^{11.7} \cdot \left(\frac{M_*}{M_{\odot}} \right)^{-1} \cdot (1+z)^{-1.5} \; [Gyr],
\end{equation}
where $M_*$ is the stellar mass at $z=z_{match/form}$, $\pm 20 \%$ random scatter assigned as a perturbation. Our prescription differs from the original one (\citealt{noeske07}) as we consider only the stellar mass 
(rather than the baryonic mass) and add a redshift-dependent correction. 

\begin{figure}[hbt!]
\includegraphics[scale=0.37]{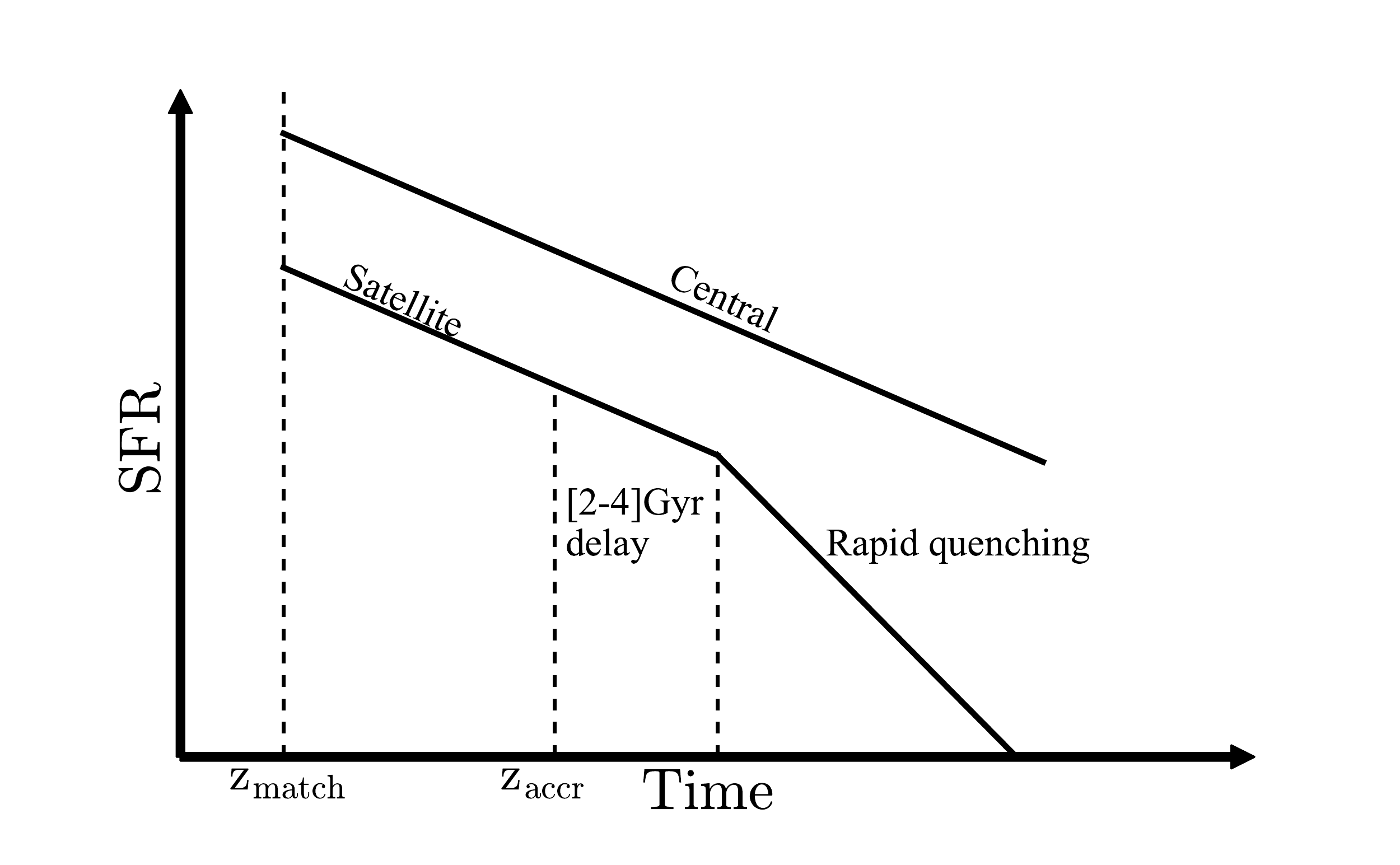}
\centering
\caption{Schematic representation of our quenching model for both centrals and satellites. The SFR of central galaxies is set at $z=z_{match/form}$ and then decays as shown by Equation \ref{eq:sfrcen}. The SFR
of satellite galaxies decays similarly to that of centrals for a "delayed" period after accretion, followed by a rapid quenching (given by $\tau_{sat}$) right after.}
\label{fig:model_scheme}
\end{figure}

The SFHs of satellites are modelled in a similar manner. Our approach is a revised version of the so-called \emph{delayed-then-rapid} quenching mode suggested by \cite{wetzel13}, where the SFRs of satellites evolve 
like those of centrals for $2-4$ Gyr after infall, and then quench rapidly according to a quenching timescale $\tau_s$. We distinguish among two kinds of satellite galaxies: satellites that were accreted before $z_{match}$, 
and those accreted after it. In the first case, the quenching timescale $\tau_s$ is assigned at $z_{match}$ by Equation \ref{eq:taus} and we assume no delayed quenching. In the second case, the quenching timescale $\tau_s$
is assigned at the redshift of accretion $z_{accr}$ and is assumed to be a random fraction $f_{\tau}$ between 0.1 and 0.5 of $\tau_c$. Hence, the SFR of satellites evolves as described by Equation \ref{eq:sfrcen} if 
$$t_{since \, infall} < t_{delay} \, ,$$ where $t_{delay}$ is randomly chosen in the range [2-4] Gyr, and as

\begin{equation} \label{eq:sfrsat}
SFR_{sat}(t)=SFR_{match/form} \cdot \exp{\left(-\frac{t}{\tau_s}\right)} 
\end{equation}
thereafter. $SFR_{match/form}$ in equations \ref{eq:sfrcen} and \ref{eq:sfrsat} is set at $z=z_{match/form}$ and derived by following Equation 2 in \cite{tomczak16}:
\begin{equation}\label{eq:tomc_par}
  \log(SFR \, [M_{\odot}/yr]) = s_0 -\log \left[ 1+\left( \frac{M_*}{M_0}\right)^{\gamma} \right]  \, ,
\end{equation}
where $s_0$ and $M_0$ are in units of $\log(M_{\odot}/yr)$ and $M_{\odot}$ respectively. As a perturbation, the model adds a random scatter in the range $\pm 0.2$ dex.
$s_0$ and $M_0$ are given by (Equation 3 in \citealt{tomczak16})
\begin{displaymath}
s_0 = 0.195  +  1.157 z  -  0.143 z^2
\end{displaymath}
\begin{equation}
\label{eq:parameterisation_all}
 \log(M_0 )= 9.244  +  0.753 z  -  0.090 z^2  
\end{equation}
\begin{displaymath}
 \gamma = -1.118 \, .
\end{displaymath}
Equation \ref{eq:tomc_par} and the set of equations \ref{eq:parameterisation_all} altogehter define the evolution with time of the SFR-$M_*$ relation with a mass-dependent slope
(for more details about the necessity of a mass-dependent slope see, e.g., \citealt{leja15,tomczak16,contini17a,contini17b}). A schematic representation of how the quenching model works is shown in Figure 
\ref{fig:model_scheme}.

This model considers both environmental and mass quenching. The environmental quenching is explicitly included in equations \ref{eq:sfrcen} and \ref{eq:sfrsat}, and it is much faster 
for satellite galaxies. The mass quenching (different from the one described in \citealt{peng10}) is implemented in the calculation of the quenching timescales, such that, for both satellite 
and central galaxies, the quenching is faster with increasing stellar mass and redshift.

\section{Results}
\label{sec:results}

\begin{figure*}
\includegraphics[scale=0.9]{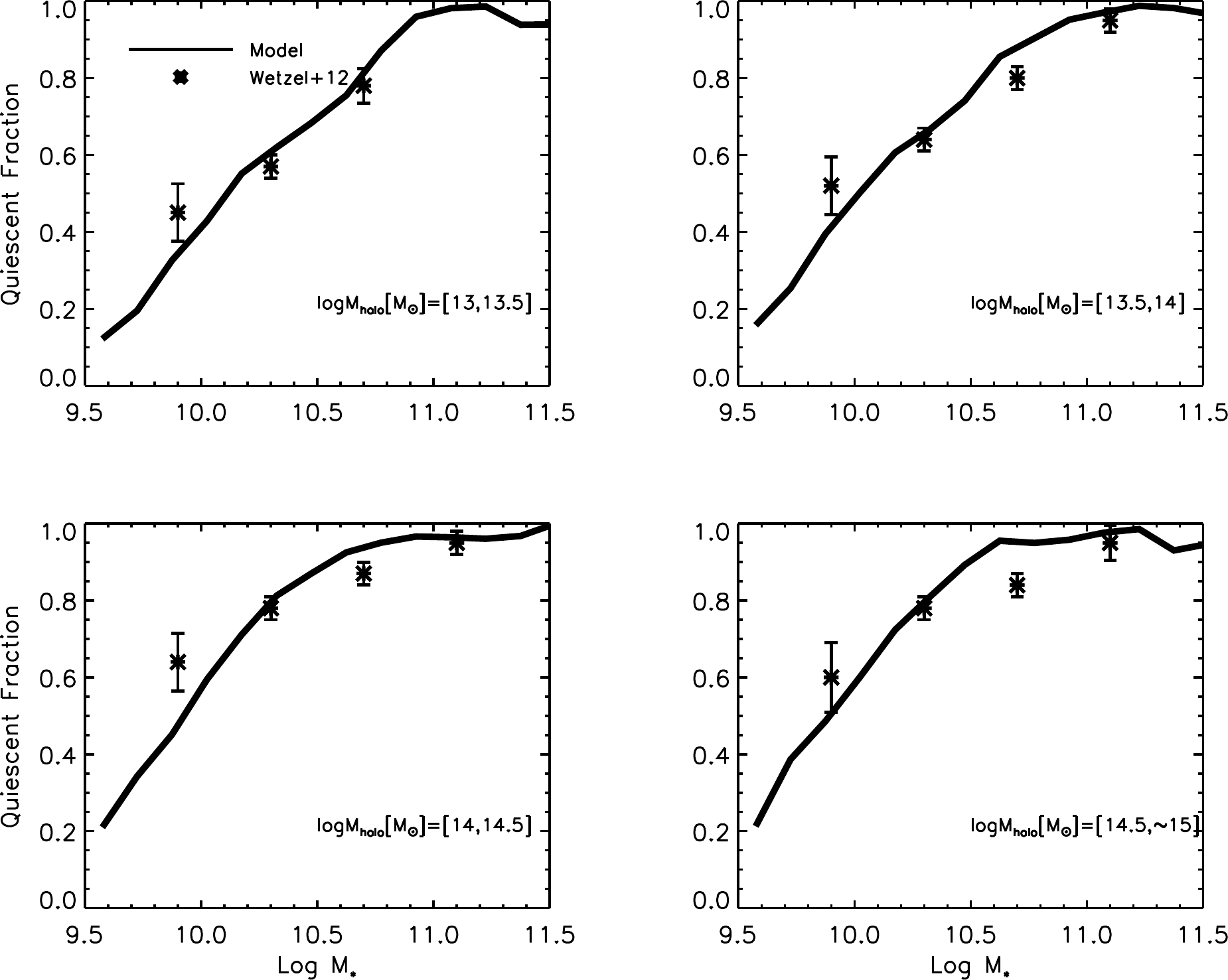}
\centering
\caption{Fraction of quiescent galaxies (black lines) as a function of stellar mass in different halo mass bins (different panels) at $z \sim 0.1$ compared with 
observed data (black crosses) by \cite{wetzel12}.}
\label{fig:quiescent_fraction}
\end{figure*}

\begin{figure*} 
\begin{center}
\begin{tabular}{cc}
\includegraphics[scale=.47]{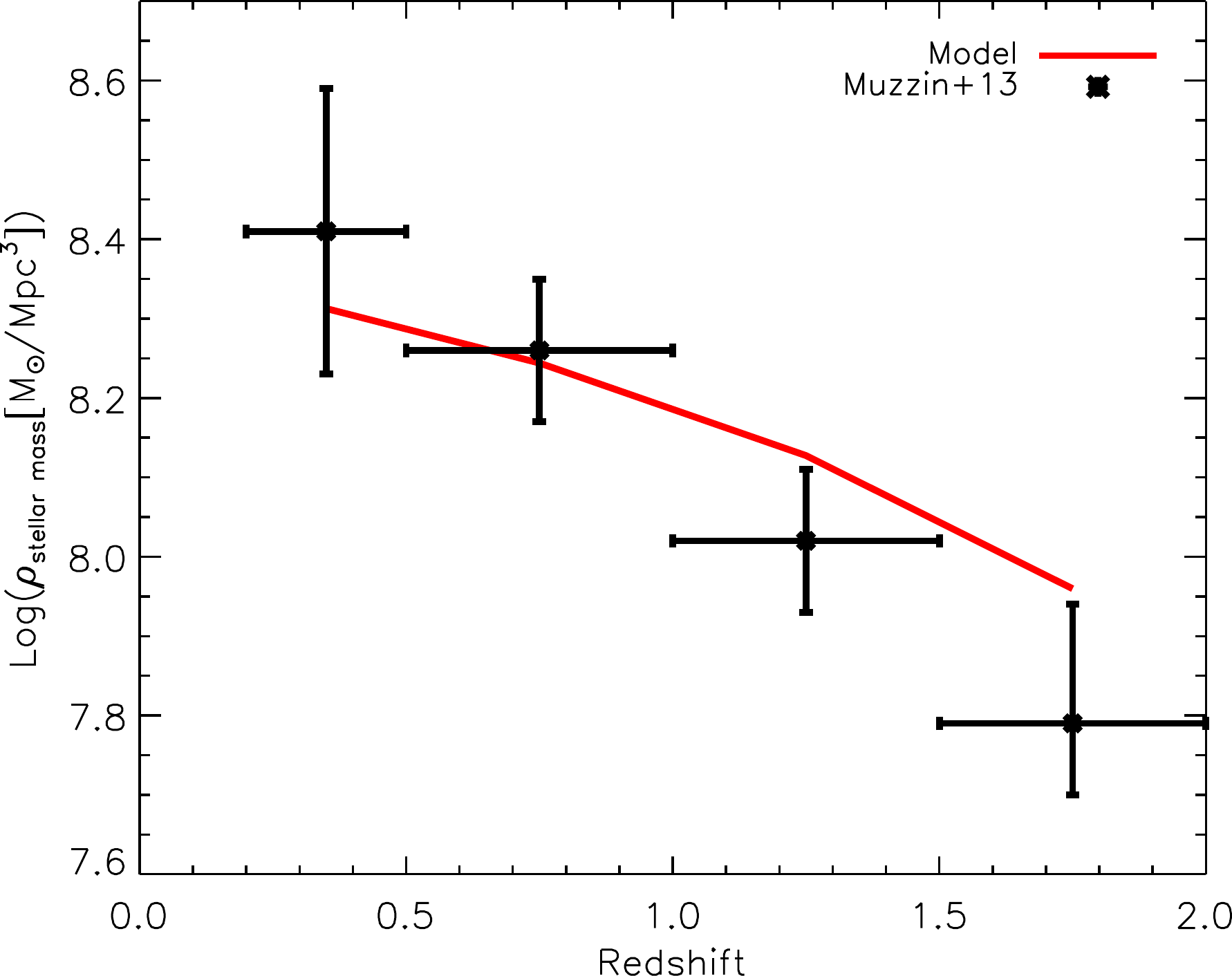} &
\includegraphics[scale=.47]{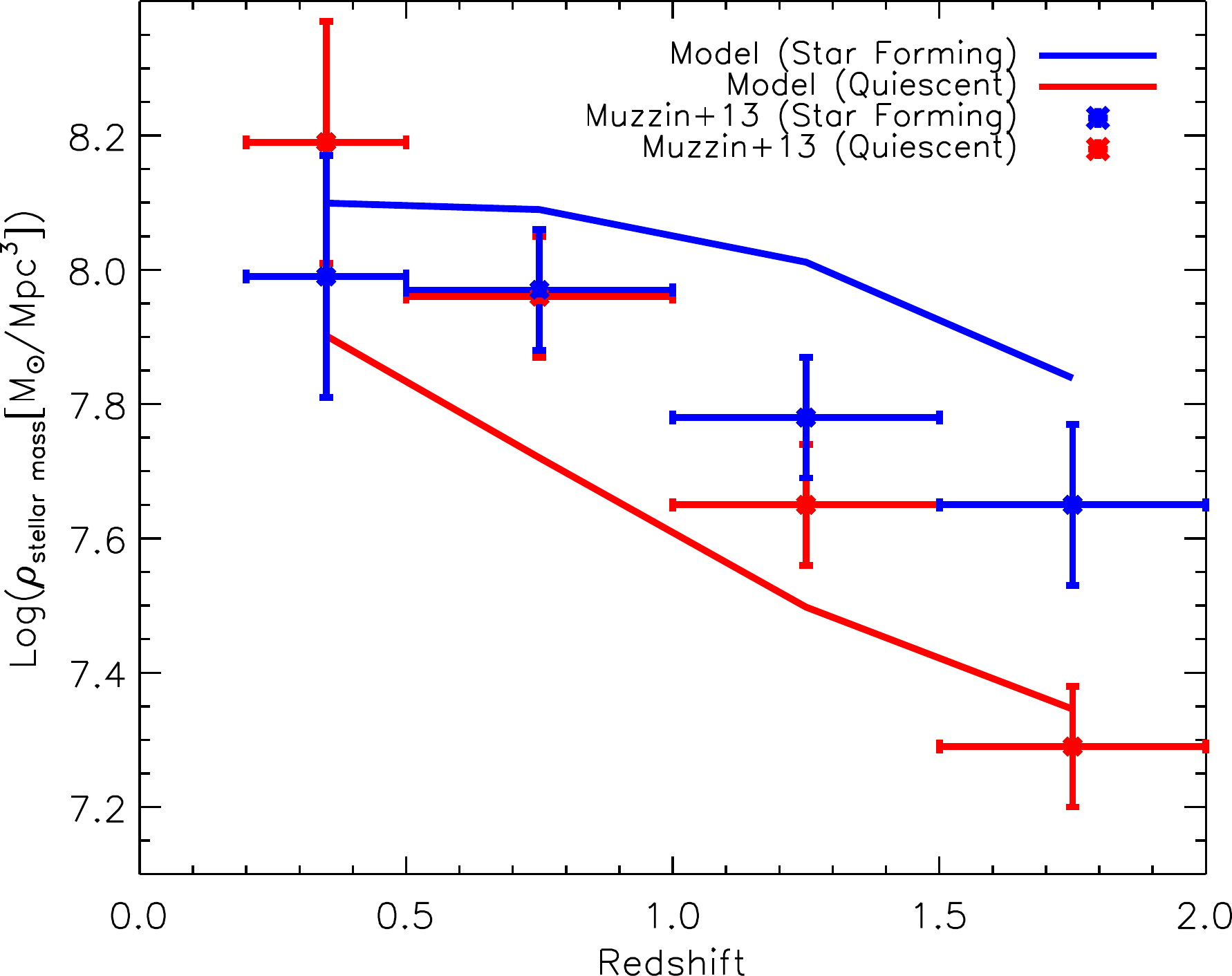} \\
\end{tabular}
\caption{Left Panel: evolution of the stellar mass density for the all population of galaxies (red line) as a function of redshift, compared with observed data
(black crosses) by \cite{muzzin13}. Right Panel: same as the left panel but for star forming (blue line and blue crosses) and quiescent (red line and red crosses)
galaxies.}
\label{fig:mass_density}
\end{center}
\end{figure*}

In this section we present our analysis and highlight the main results, which will be fully discussed in Section \ref{sec:discussion} and compared with recent studies on the same topic. All 
units are \emph{h} corrected, such that masses are expressed in $[M_{\odot}]$, SFR in $[M_{\odot}/yr]$, specific star formation rate (SSFR, which is defined as $SFR/M_*$) in $[yr^{-1}]$, and 
densities in $[M_{\odot}/Mpc^3]$.

For the purposes of our analysis we need to split the sample of galaxies in star forming and quiescent. The separation is quite arbitrary: color separation (e.g., \citealt{muzzin12,koyama13}); by using 
an offset from the star forming sequence (e.g., \citealt{trussler18,davies19}); or an SSFR cut (e.g., \citealt{wetzel12,lagana18,delucia19}). We use an SSFR cut redshift-dependent and select as star forming all 
galaxies with SSFR higher than $t_{hubble}^{-1}$, which translates in $\sim 10^{-10} yr^{-1}$ at $z=0$ (\citealt{franx08,delucia19}).

Figure \ref{fig:quiescent_fraction} shows the fraction of quiescent galaxies as a function of stellar mass as predicted by our model (solid lines), and observed data (black stars and $3\sigma$ error bars) by
\citealt{wetzel12} extracted from the SDSS Data Release 7 (\citealt{abazajian09}), for galaxies in groups/clusters of different mass as shown in the legends, at $z \sim 0.1$. For this plot only, in order to 
make a fair comparison with observed data, we use the same SSFR cut used by \cite{wetzel12}, i.e. SSFR $=10^{-11} yr^{-1}$. Our model predictions agree fairly well with the observed data in a wide range of 
halo mass, from small groups ($\log M_{halo} \sim 13$), to clusters ($\log M_{halo} \sim 15$). 

To check whether the model is also able to predict the distribution of stellar mass as a function of redshift, as it is supposed to, since it has been set to describe the evolution of the stellar mass function,
in the left panel of Figure \ref{fig:mass_density} we plot the stellar mass density as a function of redshift (red solid line), compared with observed data by \cite{muzzin13} from {\small COSMOS/UltraVISTA} survey. 
As expected, the model matches the observation within $2-\sigma$ at high redshift, and within $1-\sigma$ at $z<1$. If we plot the same quantity (right panel of Figure \ref{fig:mass_density}) for star forming (blue line and 
circles) and quiescent (red line and circles), we find a mismatch between our model and observed data such that the model underpredicts the stellar mass density of quiescent galaxies and so overpredicts that 
of star forming galaxies, independently of redshift. This is a consequence of the fact that the model is not able, according to our definition of quiescent galaxies, to predict their fraction as a function of 
redshift when compared with \cite{muzzin13} data. It must be noted that a large part of the tension can be due to the different criteria for separating the quiescent samples: a color separation in \cite{muzzin13} 
and a redshift-dependent SSFR cut in this work. This is not going to invalidate the rest of the analysis. Indeed, as pointed out by \cite{wetzel12}, color cuts can overestimate the fraction of quiescent galaxies 
because of dust reddening (see also \citealt{maller09}). In the worst case scenario, it might be that our model overestimates the SFR history of low mass galaxies (as shown by Figure \ref{fig:quiescent_fraction}), 
and this would in principle affect the environmental quenching efficiency in that stellar mass range, since low mass galaxies are generally believed to be quenched by the environment (e.g., \citealt{weisz15,fillingham16}). 
However,  even if our analysis is biased for this potential problem, we believe that the results we are going to show are robust in terms of the dependence on the environment, since the quiescent fractions of low mass 
galaxies are low at any halo mass investigated. This potential issue is going to be a key point of a forthcoming paper in preparation.

Having in mind this caveat, we proceed our analysis by going directly to the main points of the paper, i.e. the roles of mass and environment in quenching galaxies.

\subsection{Environmental Quenching}
\label{sec:env_quenching}

\begin{figure*} 
\begin{center}
\begin{tabular}{cc}
\includegraphics[scale=.5]{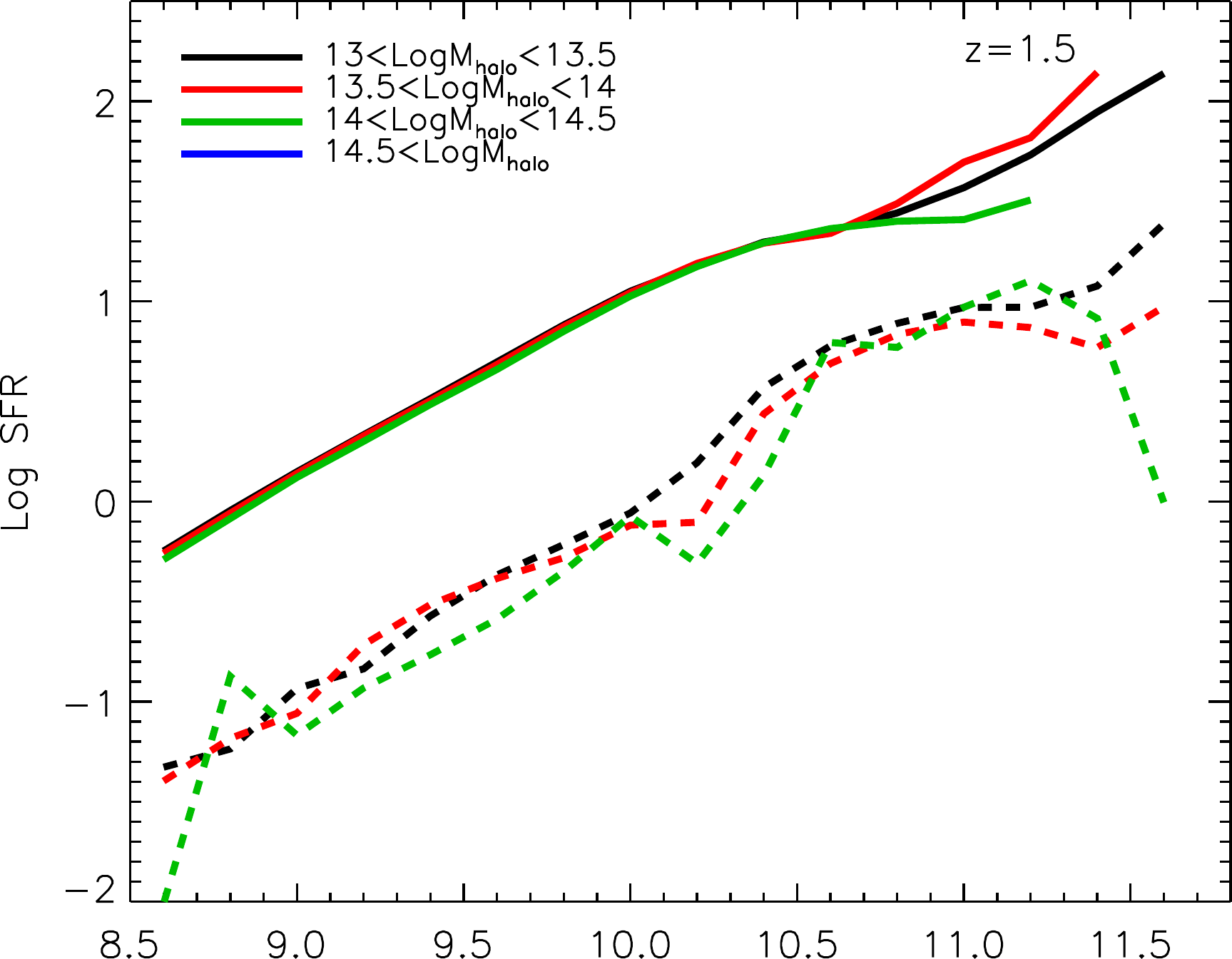} &
\includegraphics[scale=.5]{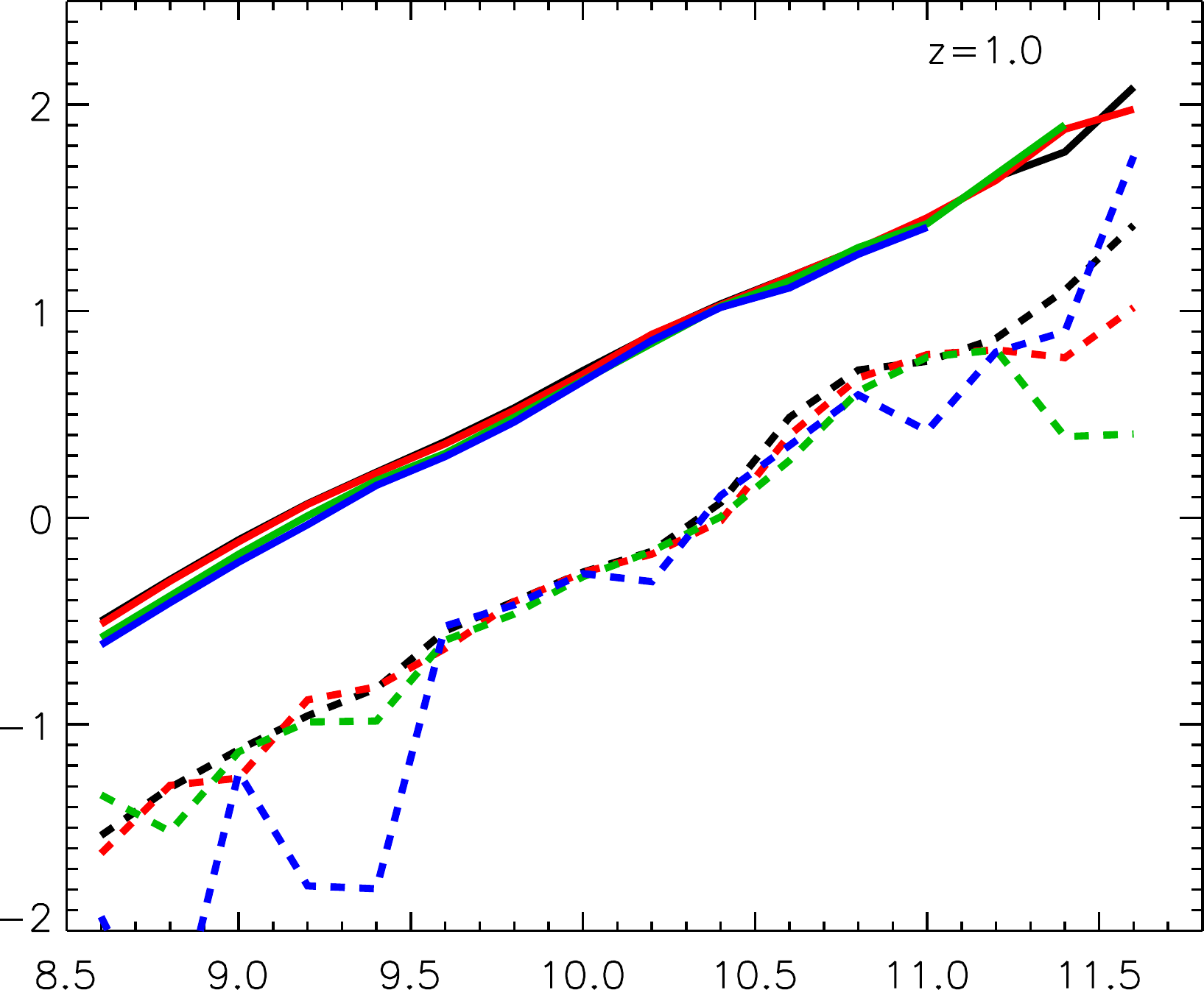} \\
\includegraphics[scale=.5]{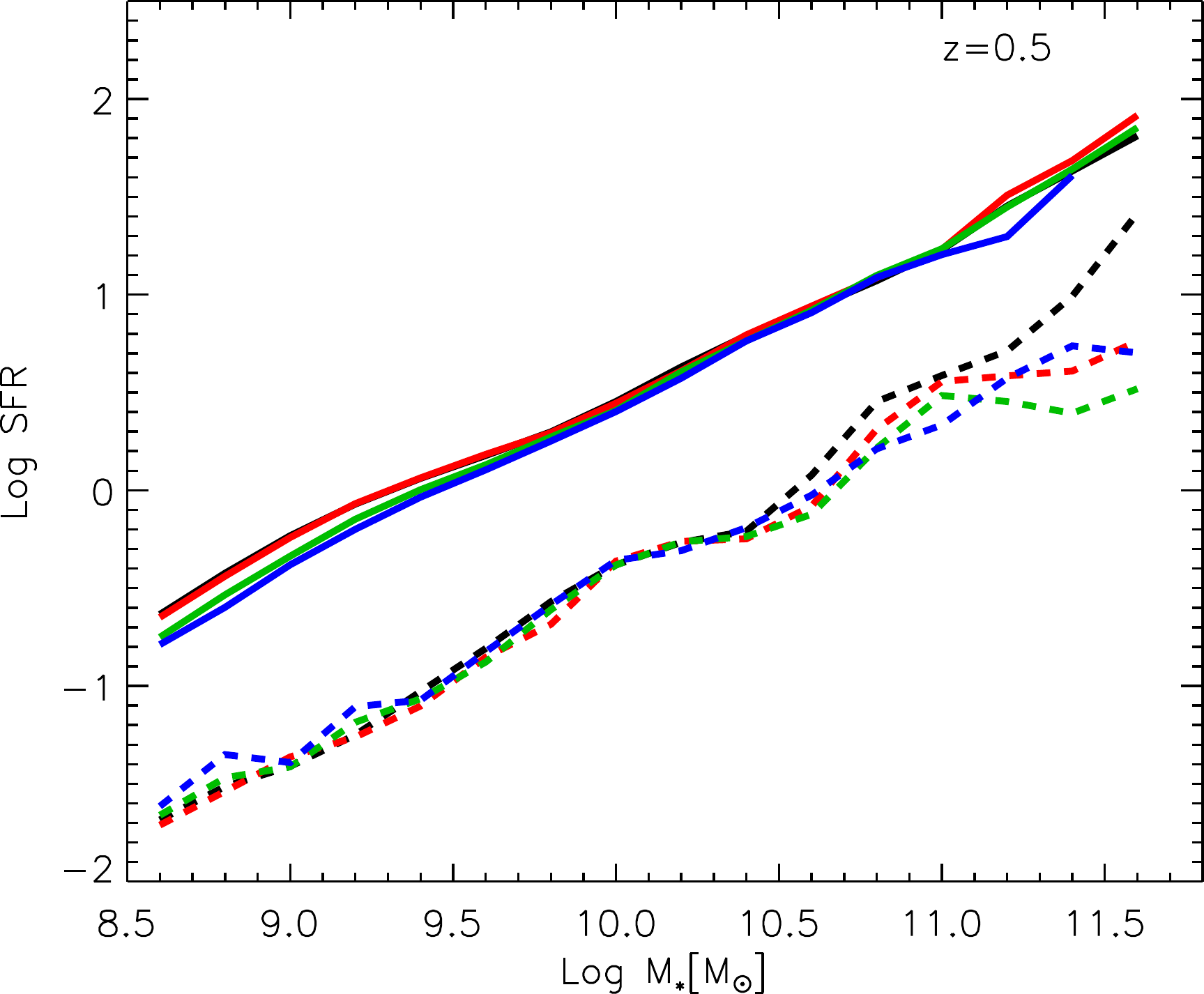} &
\includegraphics[scale=.5]{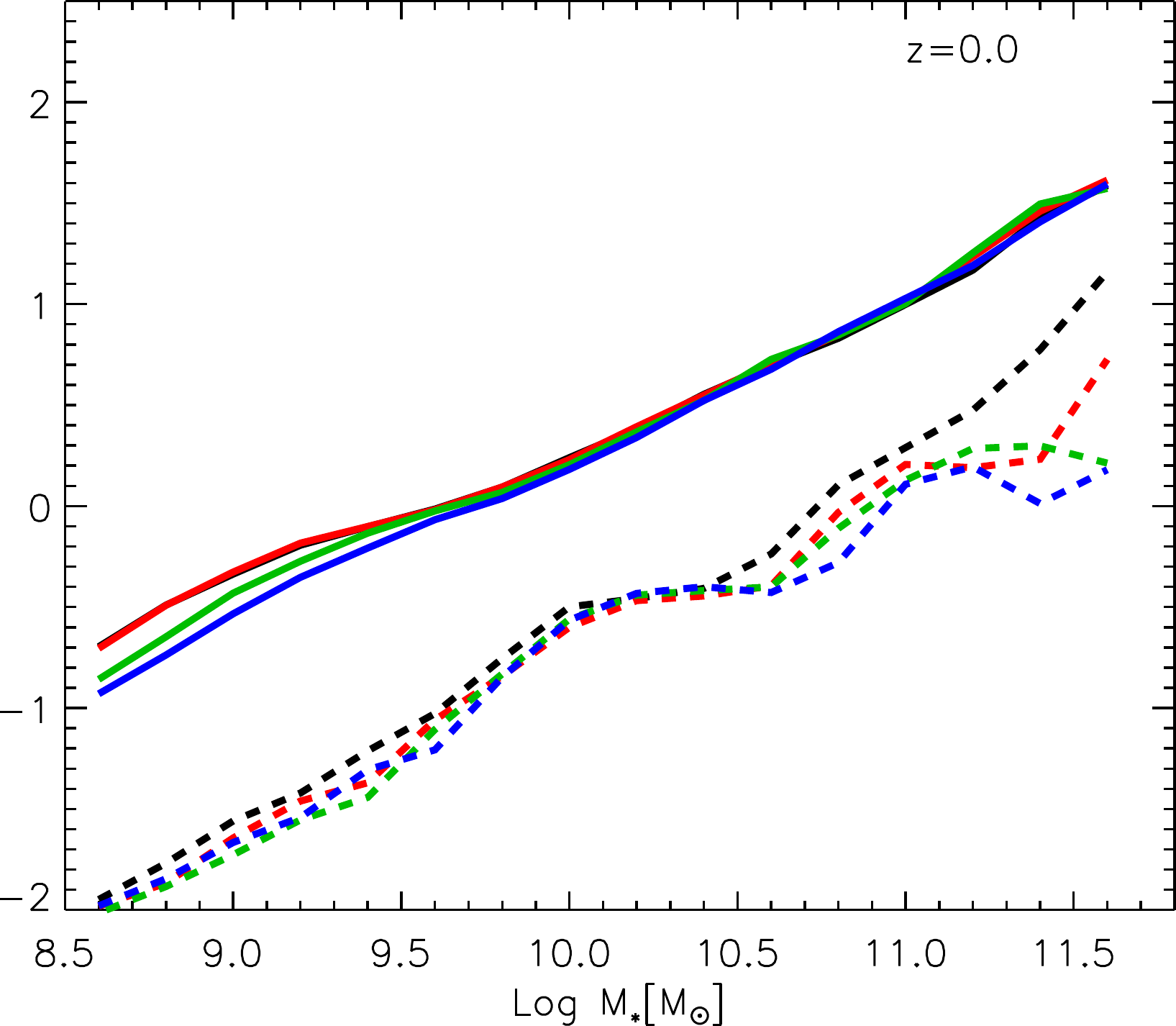} \\
\end{tabular}
\caption{Evolution of the star formation rate - stellar mass relation as a function of redshift (different panels) for star forming (solid lines) and 
quiescent (dashed lines) galaxies residing in haloes of different mass, as indicated in the legend. The average $1\sigma$ scatter is around 0.2/0.25 dex 
for star forming/quiescent galaxies, mostly independent on redshift and on halo mass. Clearly, the environment does not play any role in the SFR-$M_*$ 
relation for both star forming and quiescent galaxies, at any redshift investigated.}
\label{fig:sfrmass_qs_halomass}
\end{center}
\end{figure*}

\begin{figure*} 
\begin{center}
\begin{tabular}{cc}
\includegraphics[scale=.48]{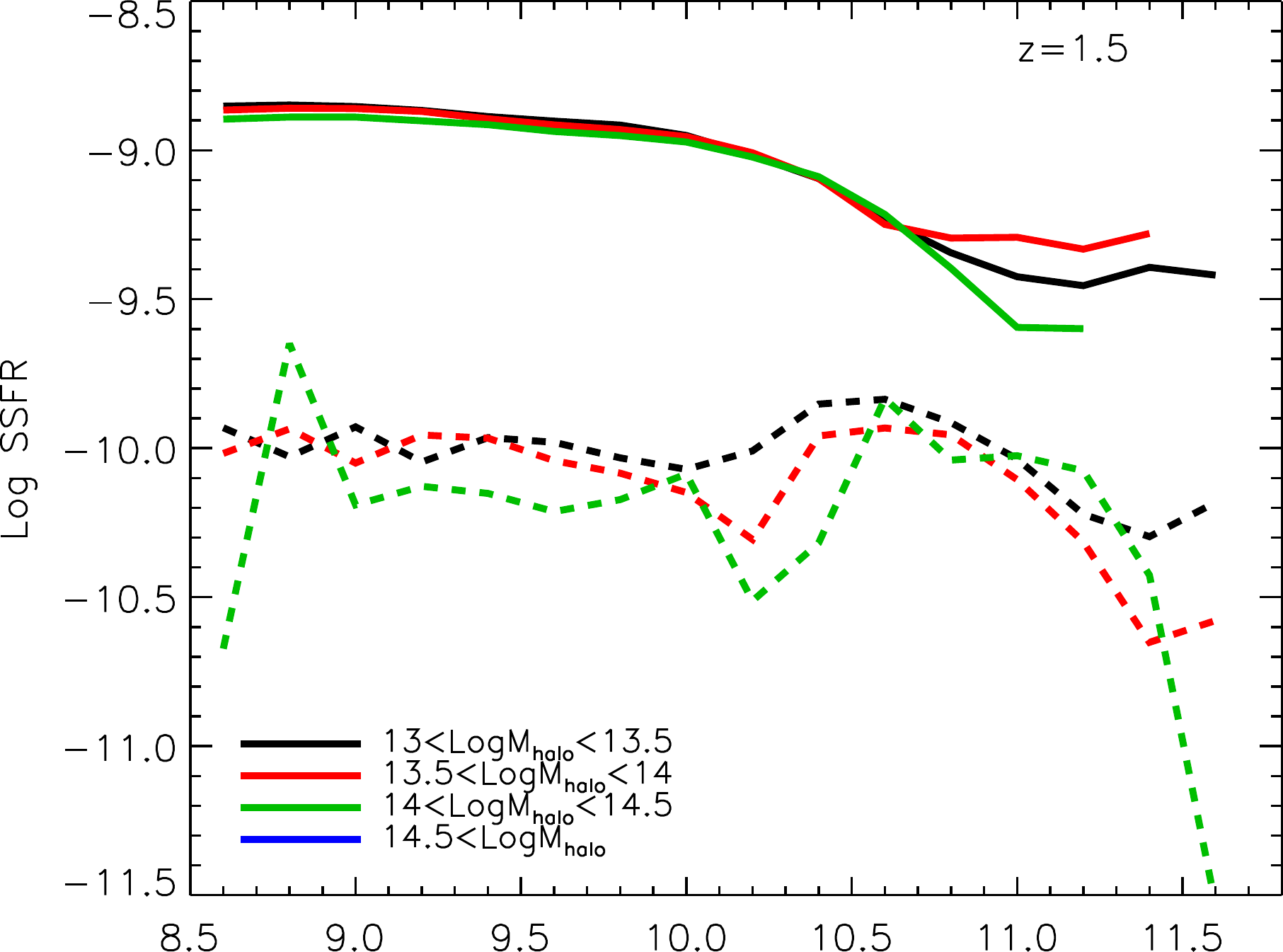} &
\includegraphics[scale=.48]{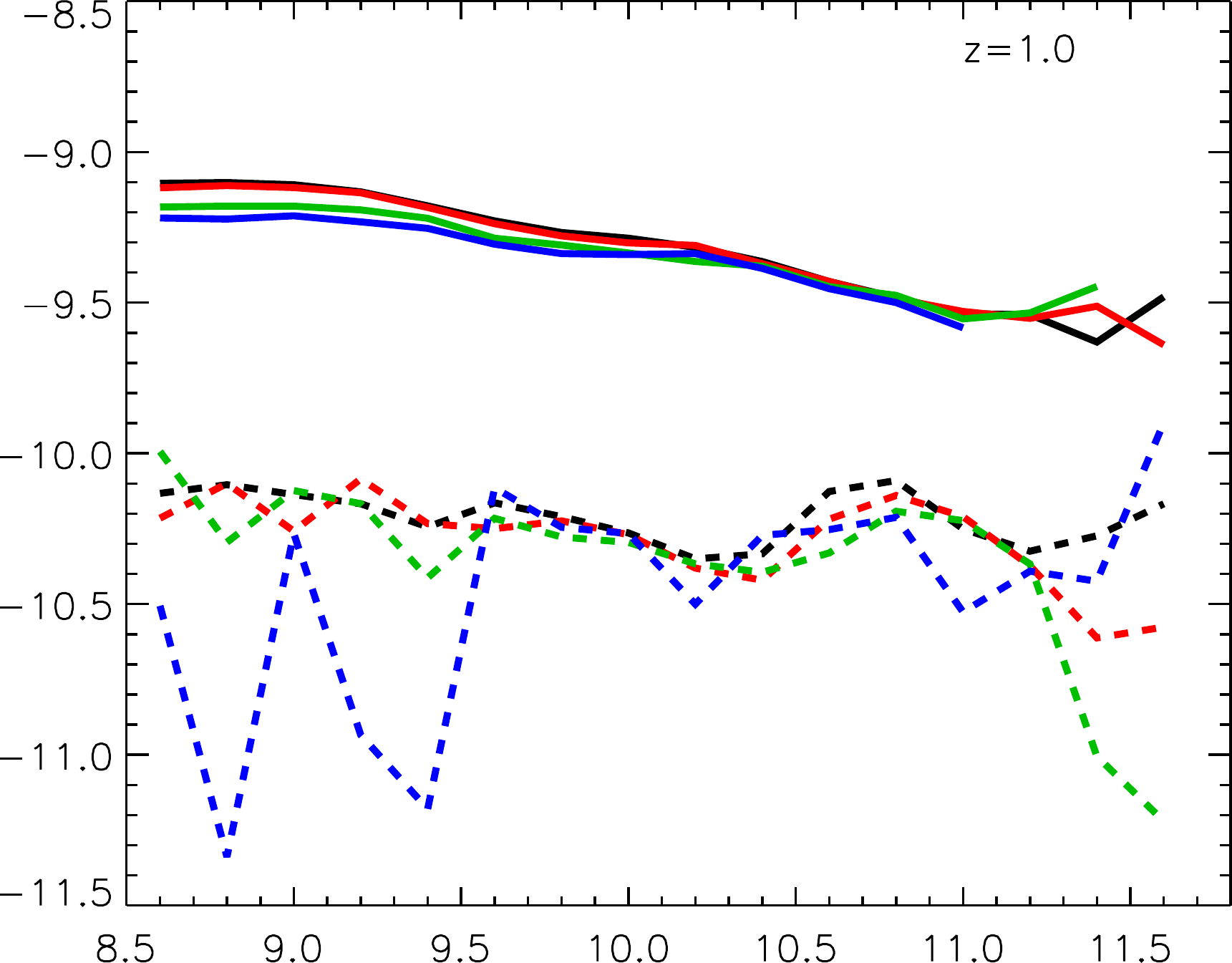} \\
\includegraphics[scale=.48]{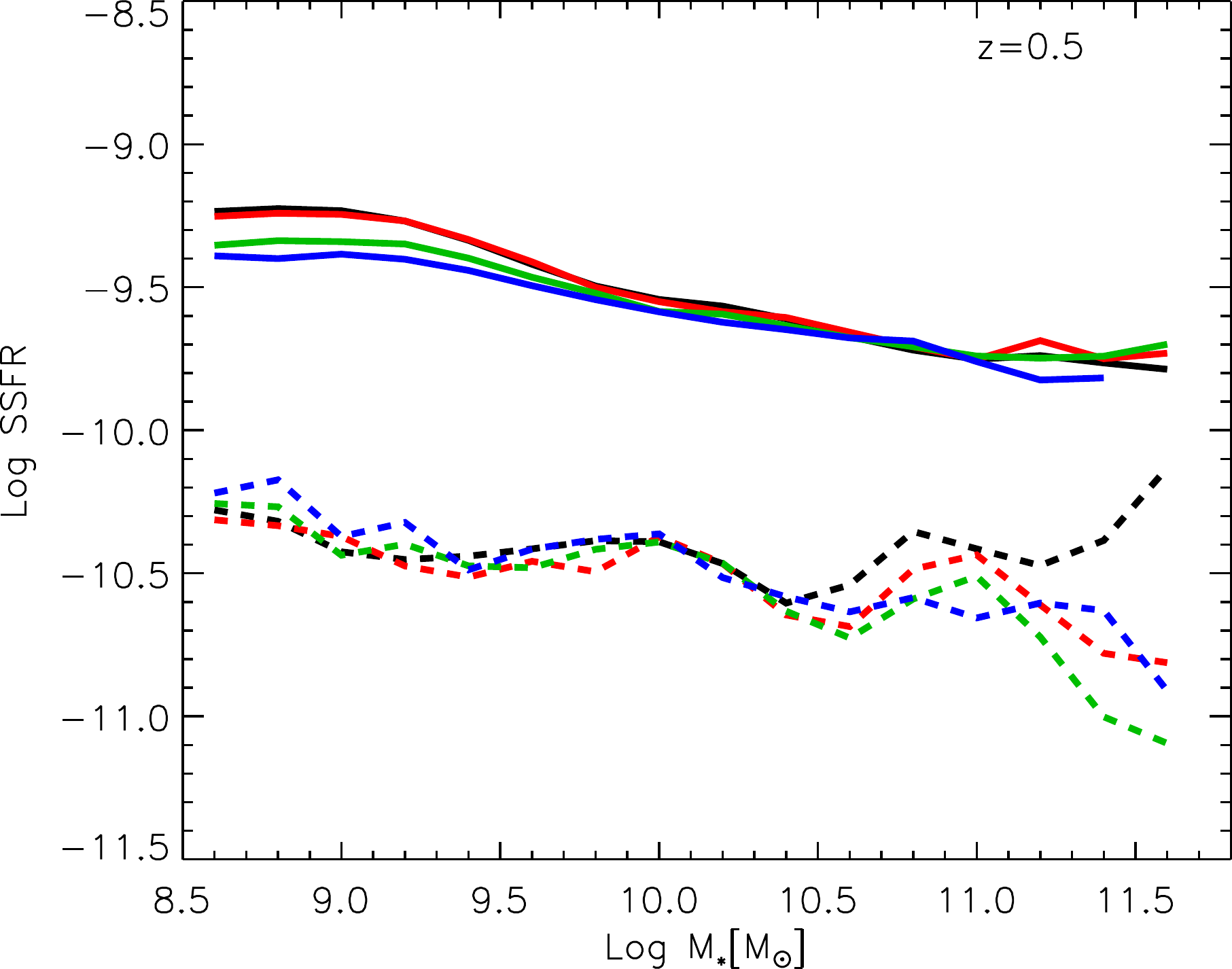} &
\includegraphics[scale=.48]{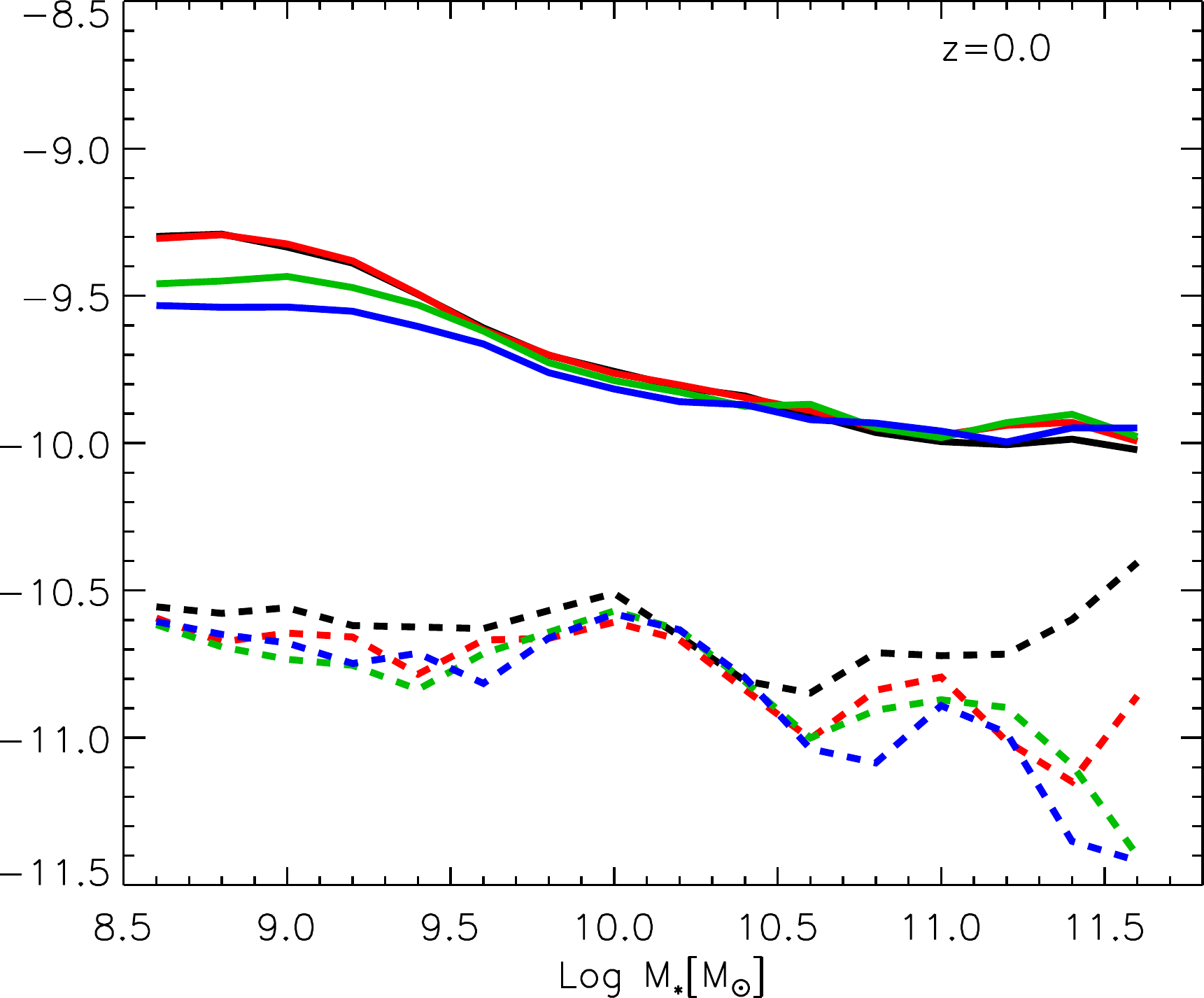} \\
\end{tabular}
\caption{Evolution of the specific star formation rate - stellar mass relation as a function of redshift (different panels) for star forming (solid lines) and 
quiescent (dashed lines) galaxies residing in haloes of different mass, as indicated in the legend. The average $1\sigma$ scatter is around 0.15/0.2 dex
for star forming/quiescent galaxies, mostly independent on redshift and on halo mass. As for the SFR shown in Figure \ref{fig:sfrmass_qs_halomass}, the 
SSFR at fixed stellar mass is independent of environment, for both star forming and quiescent galaxies.}
\label{fig:ssfrmass_qs_halomass}
\end{center}
\end{figure*}

It has been pointed out by several authors (e.g., \citealt{wetzel12,muzzin12,koyama13,darvish16,lagana18}) that, in order to extract the dependence of the SFR (and so SSFR) on stellar mass/environment, one has 
to study the quantity at fixed environment/stellar mass. In Figure \ref{fig:sfrmass_qs_halomass} we focus on the role of the environment in shaping the SFR of galaxies by plotting the SFR-$M_*$ relation for galaxies 
residing in clusters of different mass (different colors), for star forming (solid lines) and quiescent (dashed lines) galaxies at different redshifts (different panels). As clearly shown by the plots, the environment 
(defined here as the halo mass in which galaxies reside) does not play any role in the SFR-$M_*$ relation for both star forming and quiescent galaxies, at any redshift investigated. This result is in perfect agreement 
with other studies, e.g., with \cite{koyama13}, who studied the environmental dependence of the SFR-$M_*$ relation for star forming galaxies since $z \sim 2$ with H$_{\alpha}$ emitters in clusters and field environments. 
They conclude that such relation for star forming galaxies is environment independent at any epoch, even considering dust attenuation. We support their results and extend the same conclusion to quiescent galaxies, 
although some environmental dependence is seen for very massive quiescent galaxies.

Figure \ref{fig:ssfrmass_qs_halomass} shows the same information shown by Figure \ref{fig:sfrmass_qs_halomass}, but for the SSFR. As for the SFR, the SSFR at fixed stellar mass is independent of environment, for both
star forming and quiescent galaxies (although again, the very massive quiescent galaxies seem to show some dependence). Our results agree well with former studies (e.g., \citealt{muzzin12,lagana18}). 
\cite{muzzin12} studied the effects of stellar mass and environment on the SFR and SSFR of galaxies in the redshift range $0.8<z<1.2$ for a spectroscopic selected sample of galaxies in clusters and field extracted 
from the Gemini Cluster Astrophysics Spectroscopic Survey. They find that, once the SSFR is plotted at fixed stellar mass, it is environment independent. It is worth noting that, however, their definition of environment 
is the clustercentric distance, rather than halo mass. Moreover, for the least massive ($\log M_* \lesssim 9.3$) star forming galaxies, there seems to be a trend with decreasing redshift for which both the SFR and SSFR 
decrease with increasing halo mass, which might be a hint of environment dependence at least in that stellar mass range. However, we note that the average difference in that stellar mass range between the two extreme 
halo mass bins is less than 0.2 dex, i.e. within the typical SFR dispersion around the main sequence of star forming galaxies.

\subsection{Mass Quenching}
\label{sec:mass-quenching}

\begin{figure*} 
\begin{center}
\begin{tabular}{cc}
\includegraphics[scale=.5]{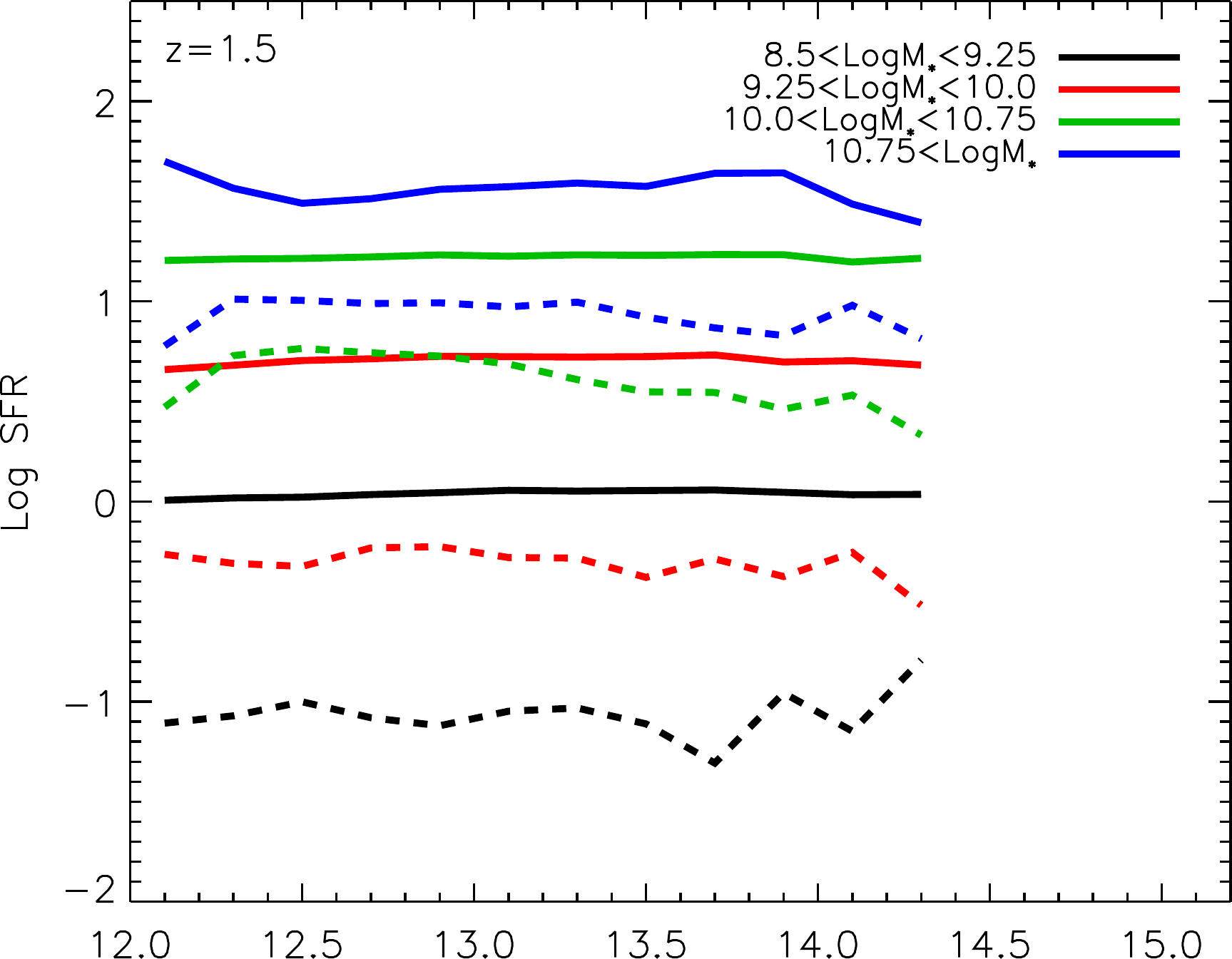} &
\includegraphics[scale=.5]{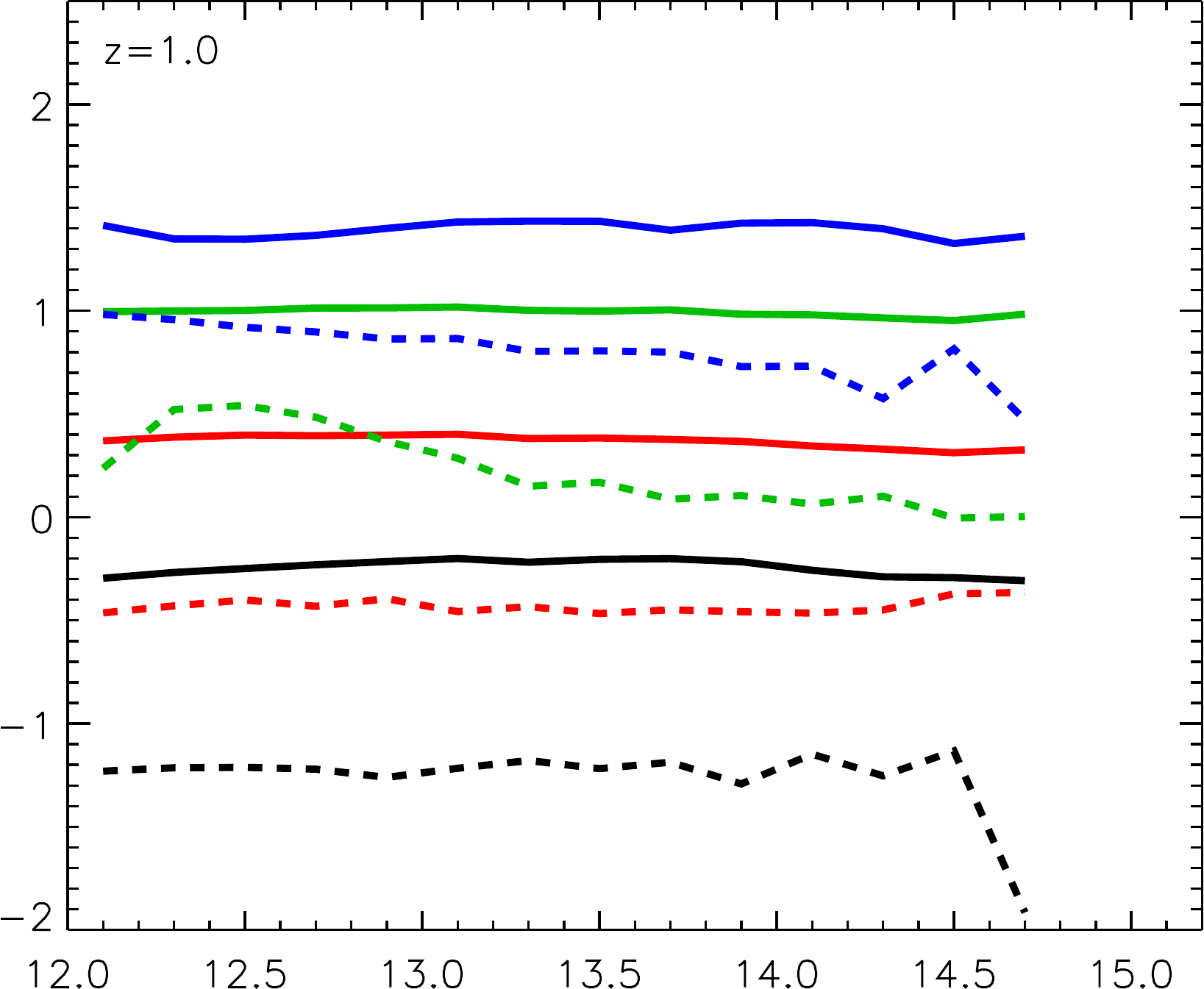} \\
\includegraphics[scale=.5]{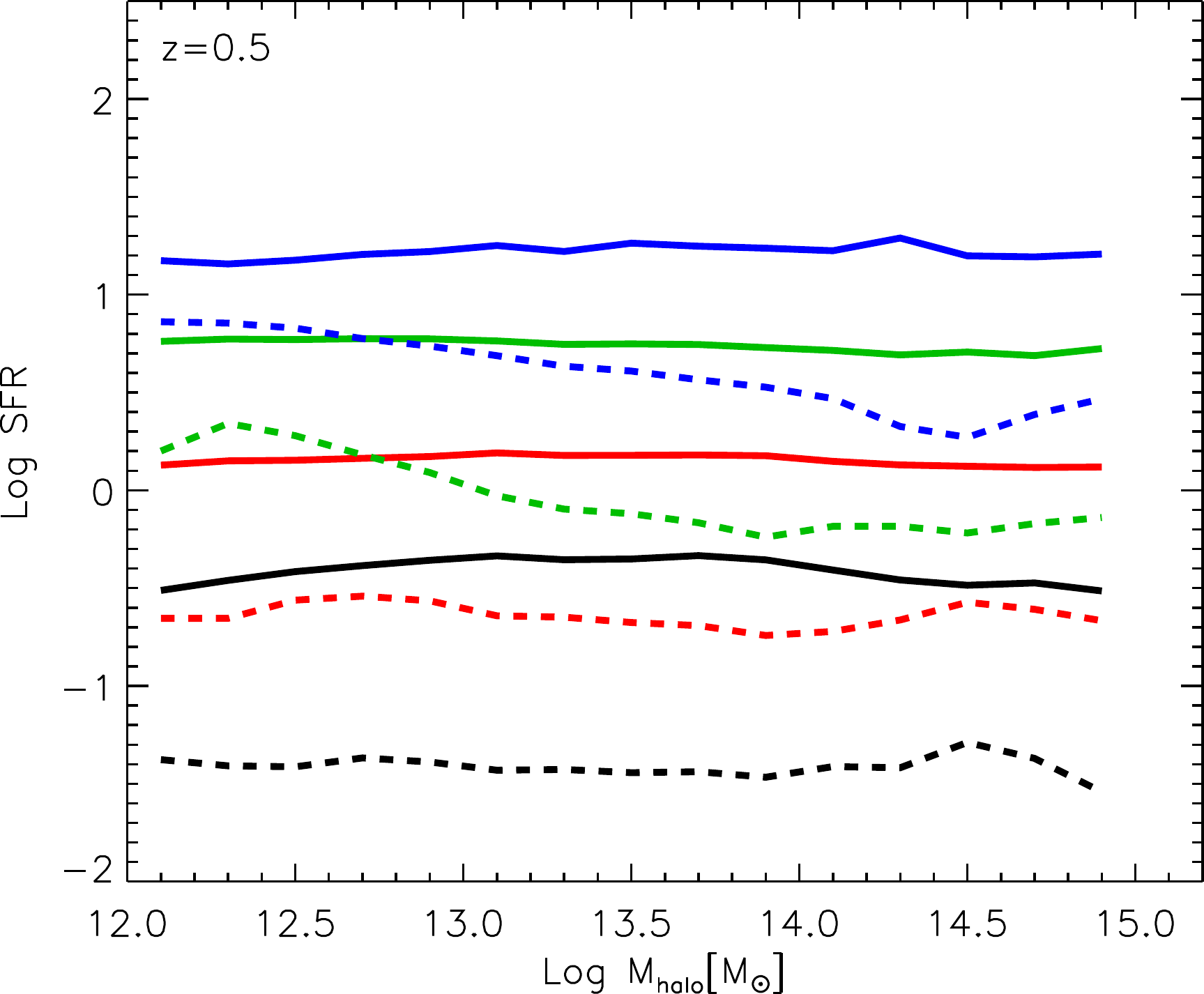} &
\includegraphics[scale=.5]{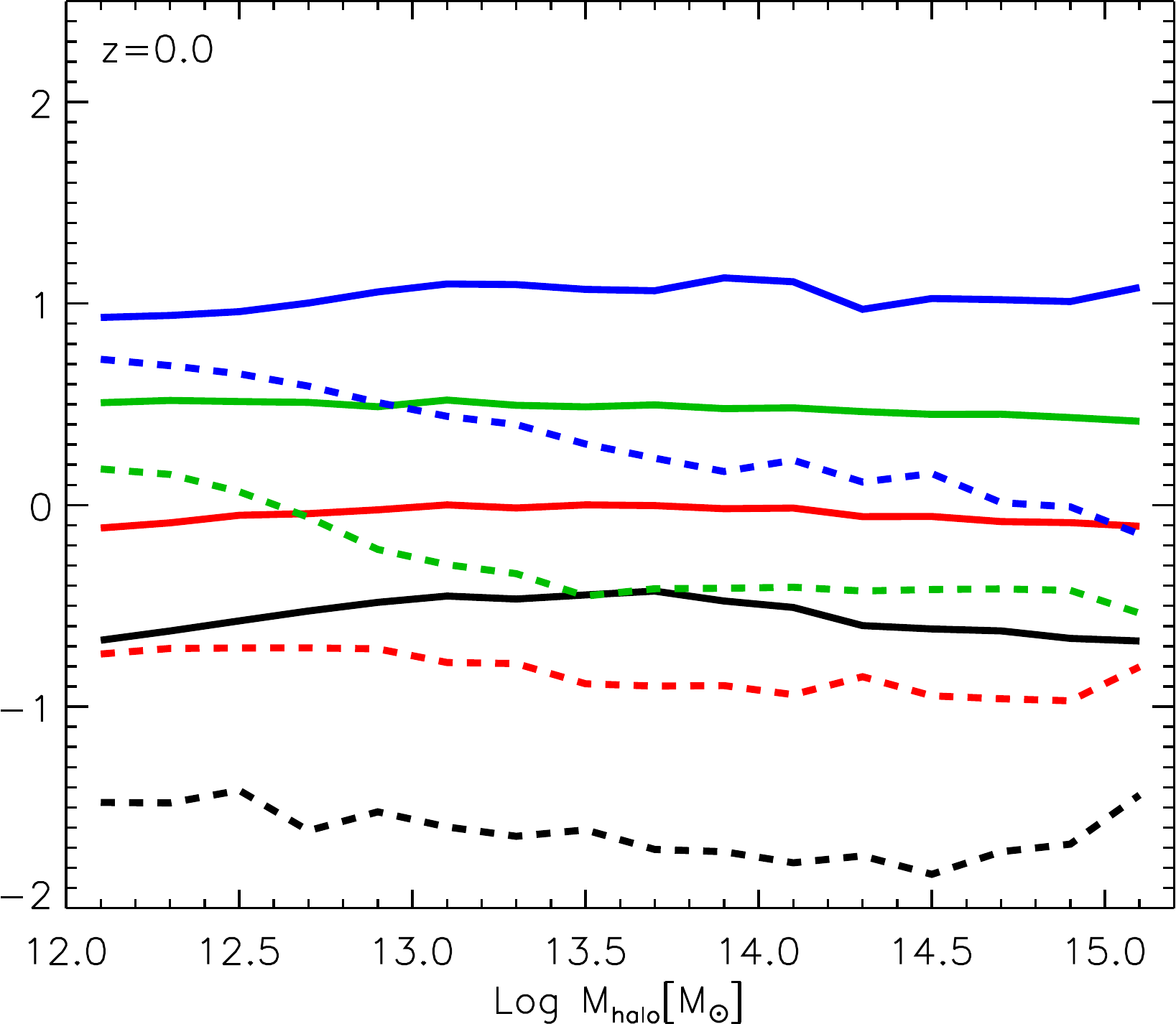} \\
\end{tabular}
\caption{Evolution of the star formation rate - halo mass relation as a function of redshift (different panels) for star forming (solid lines) and 
quiescent (dashed lines) galaxies in different stellar mass bins, as indicated in the legend. The average $1\sigma$ scatter is around 0.15/0.2 dex, 
for star forming/quiescent galaxies, independently on redshift and stellar mass. The SFR is independent of halo mass (as found in Figure 
\ref{fig:sfrmass_qs_halomass}) at any redshift, but at a given halo mass, the SFR is stronlgy dependent on stellar mass, for both star forming and 
quiescent galaxies. This is a clear evidence of mass quenching.}
\label{fig:sfr_halomass_qs_mass}
\end{center}
\end{figure*}

\begin{figure*} 
\begin{center}
\begin{tabular}{cc}
\includegraphics[scale=.48]{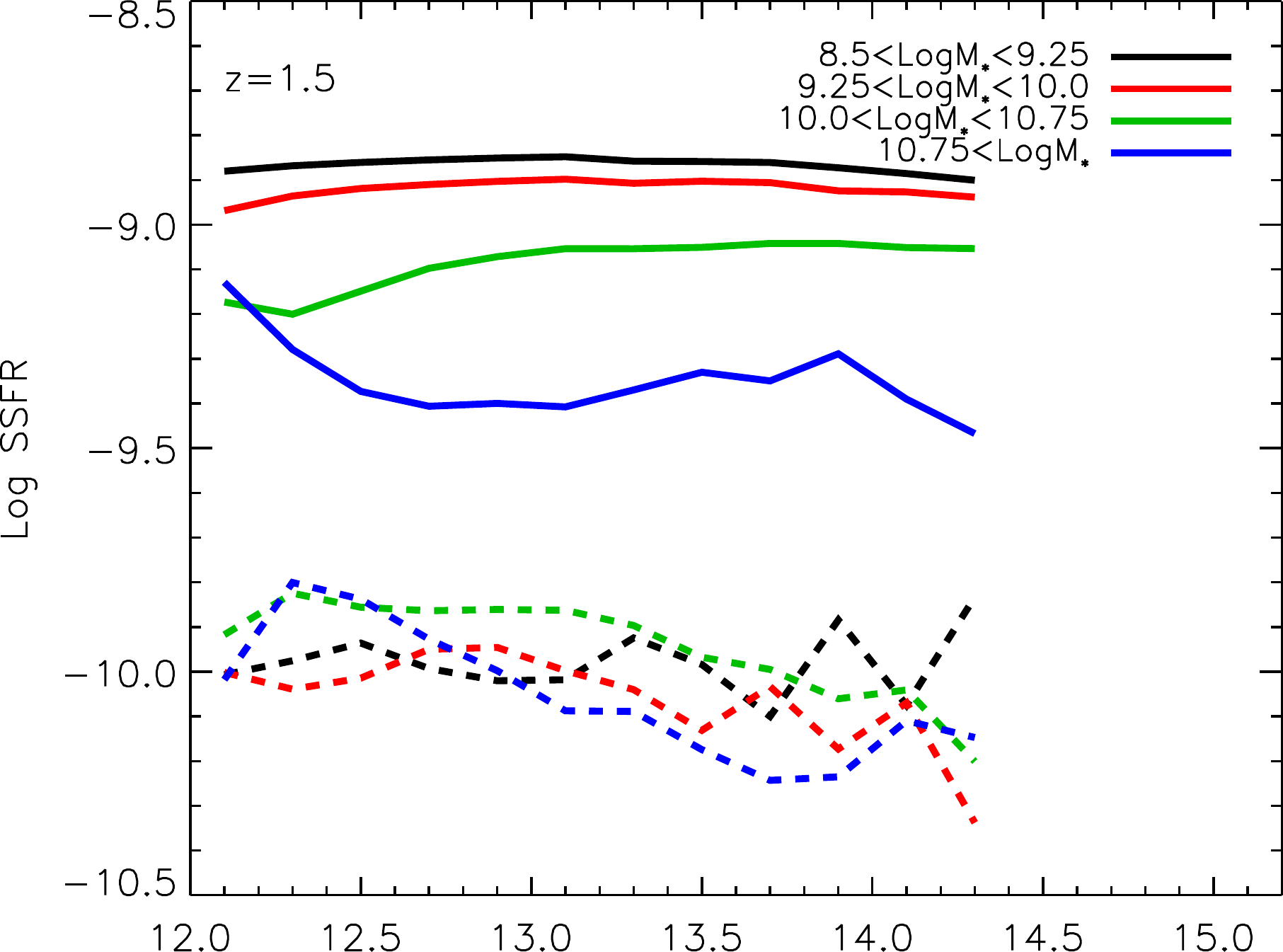} &
\includegraphics[scale=.48]{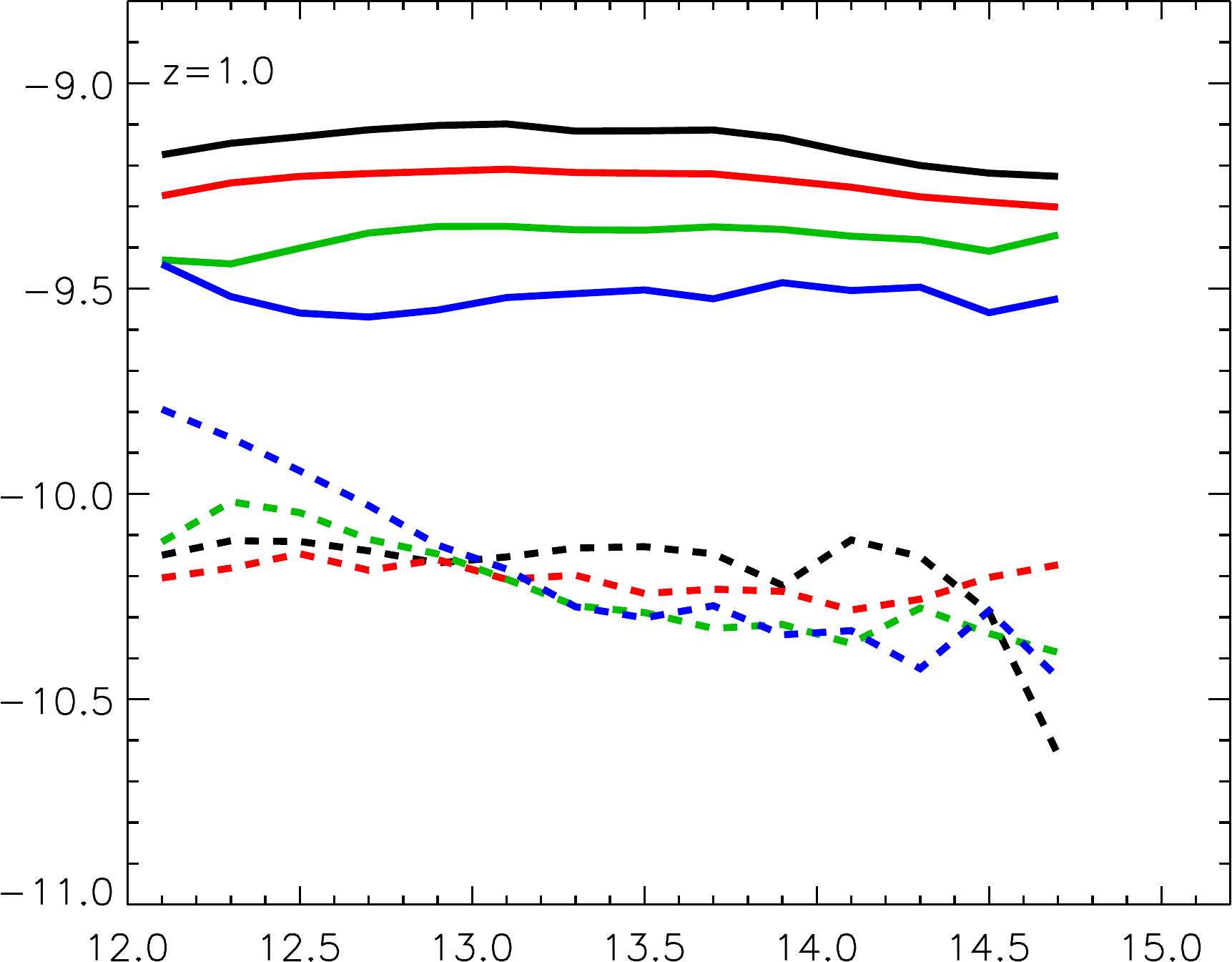} \\
\includegraphics[scale=.48]{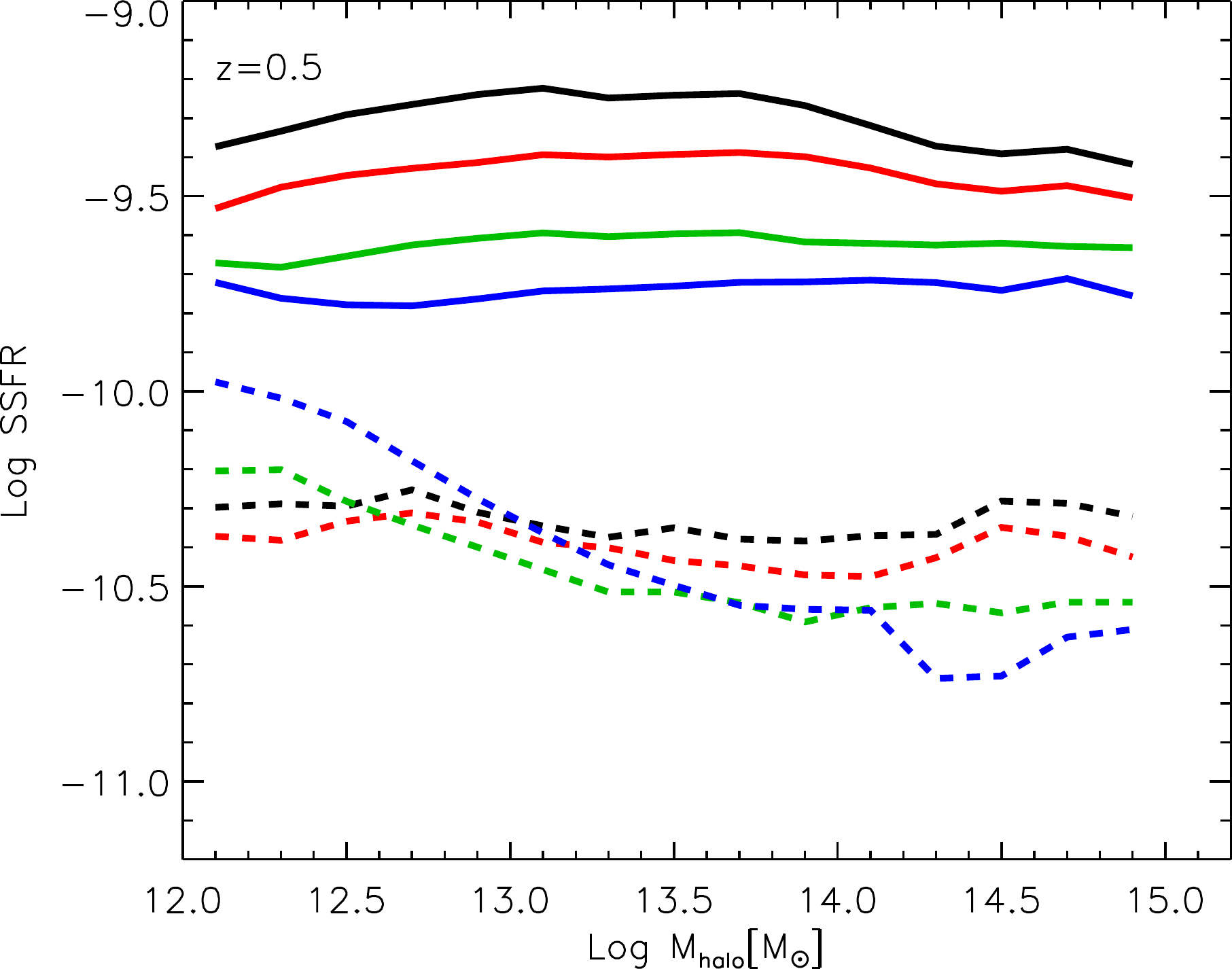} &
\includegraphics[scale=.48]{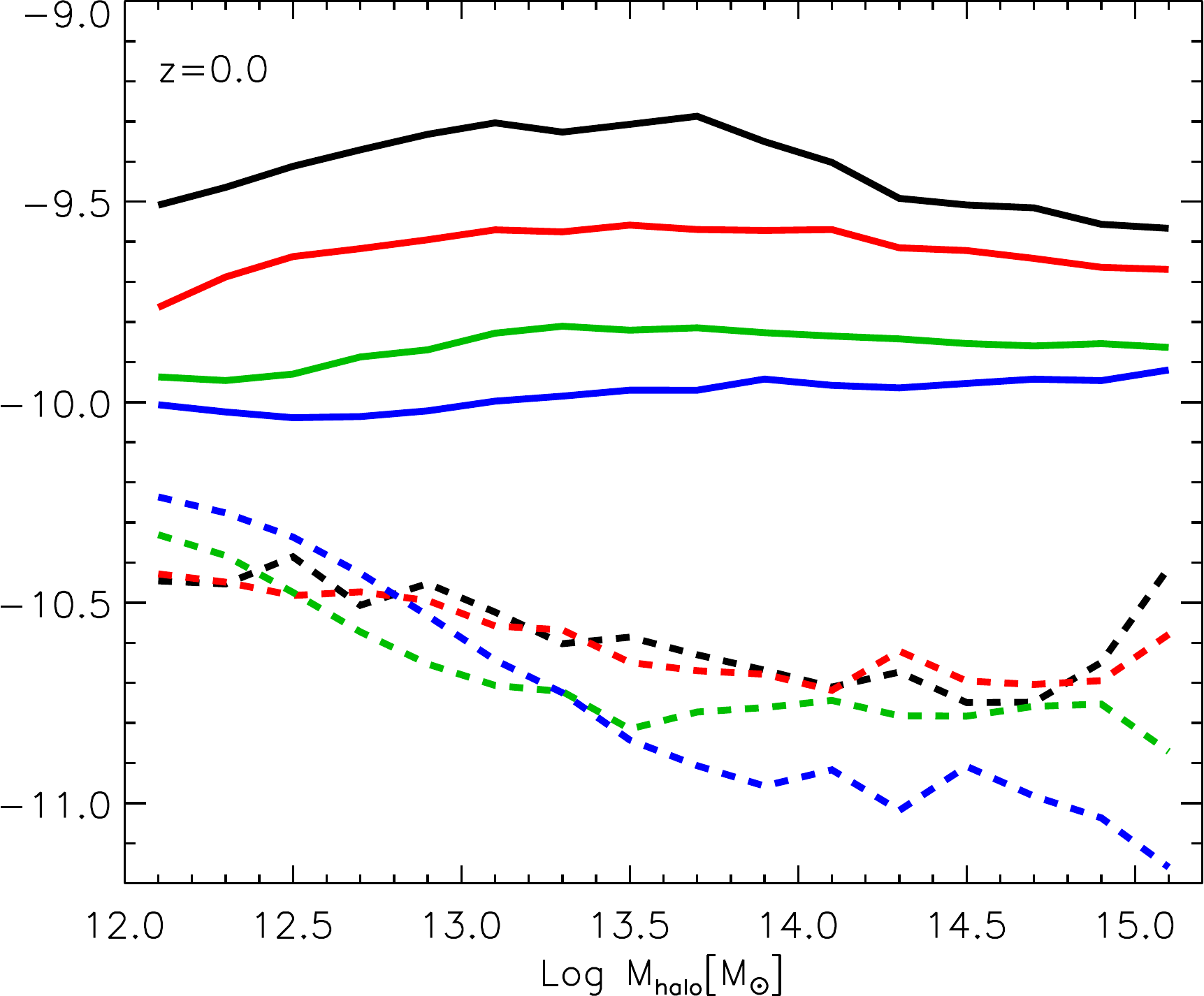} \\
\end{tabular}
\caption{Evolution of the specific star formation rate - halo mass relation as a function of redshift (different panels) for star forming (solid lines) and 
quiescent (dashed lines) galaxies in different stellar mass bins, as indicated in the legend. The average $1\sigma$ scatter is around 0.1/0.15 dex, 
for star forming/quiescent galaxies, independently on redshift and stellar mass. Similarly to Figure \ref{fig:sfr_halomass_qs_mass}, the same trend 
is found for the SSFR of the star forming sample, while the trend does not appear clear for the quiescent one.}
\label{fig:ssfr_halomass_qs_mass}
\end{center}
\end{figure*}

\begin{figure*} 
\begin{center}
\begin{tabular}{cc}
\includegraphics[scale=.47]{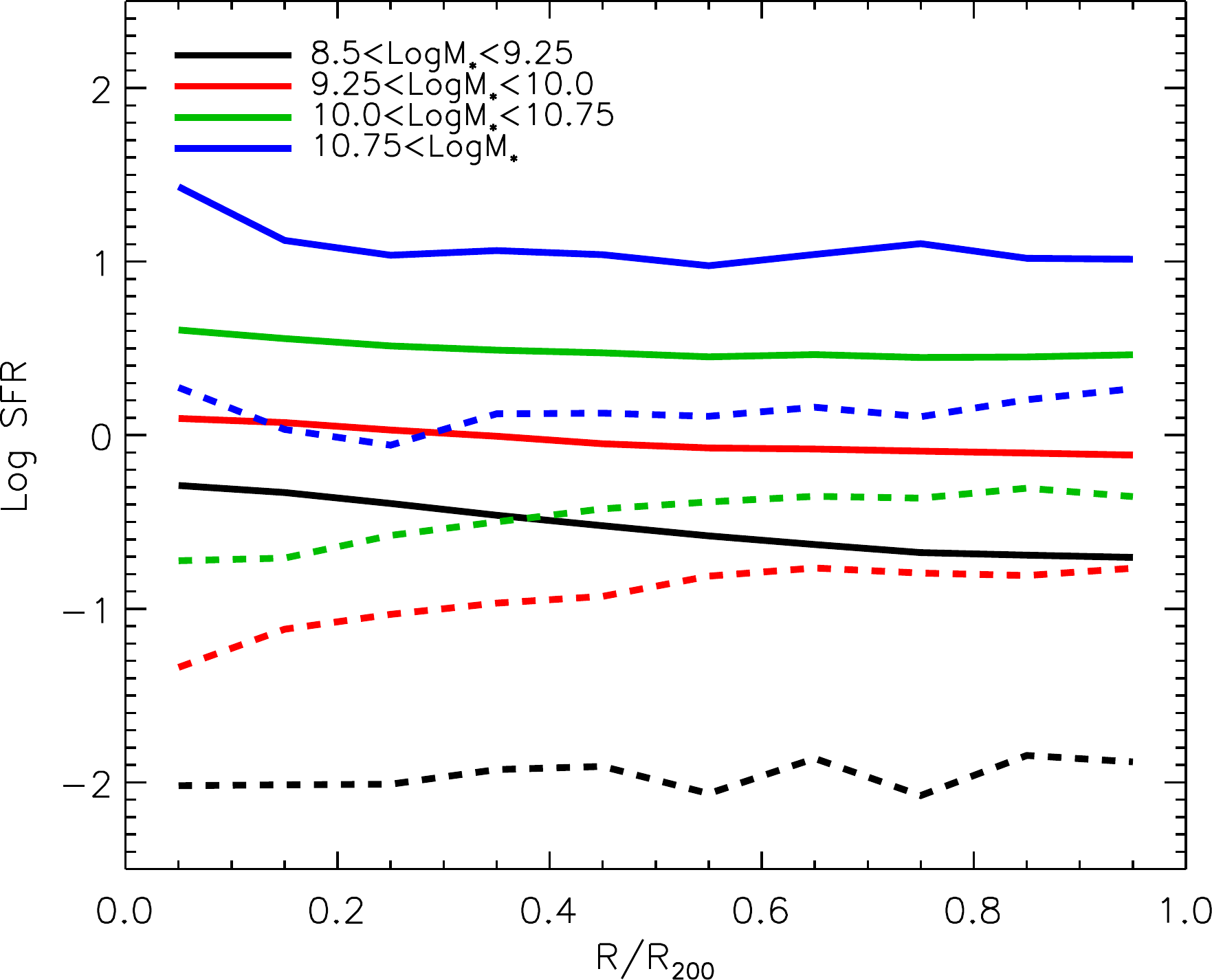} &
\includegraphics[scale=.47]{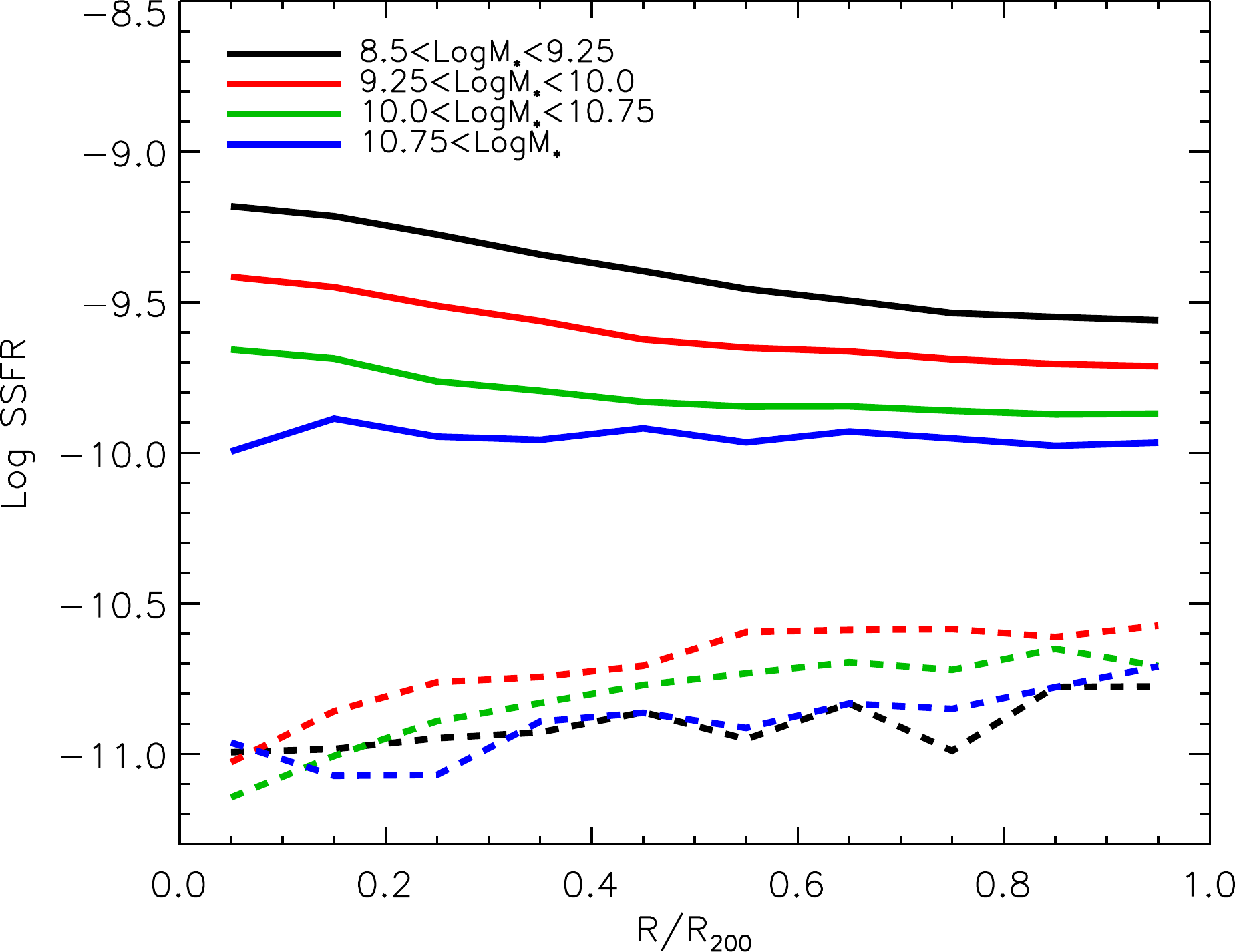} \\
\end{tabular}
\caption{Star formation rate (left panel) and specific tar formation rate (right panel) of star forming (solid lines) and quiescent (dashed lines) galaxies as a function of distance from the halo center 
for galaxies in different stellar mass bins, at $z=0$. The average $1\sigma$ scatter is around 0.2/0.25 dex (left panel), and 0.15/0.2 (right panel), for star forming/quiescent galaxies, independently 
on stellar mass. Altogether, considering the results found above, this strongly suggests that at fixed environmental conditions, may be them given by the typical halo mass or clustercentric distance, 
the SFR and SSFR depend on stellar mass.}
\label{fig:distance_mass}
\end{center}
\end{figure*}

We now move the subject of the analysis to the role of mass quenching, i.e. we study the SFR and SSFR as a function of environment (defined as halo mass) at fixed stellar mass. 
Figure \ref{fig:sfr_halomass_qs_mass} shows the SFR of star forming (solid lines) and quiescent (dashed lines) galaxies as a function of halo mass, for galaxies in different stellar mass bins as indicated 
in the legend, and at different redshifts (different panels). The SFR is independent of halo mass (as found above) at any redshift, and the interesting feature is that, at a given halo mass, the SFR is 
stronlgy dependent on stellar mass, for both star forming and quiescent galaxies. Indeed, in a range of three orders of magnitude in halo mass, the average difference between the SFR of the least massive 
stellar mass range ($8.5<\log M_* <9.25$) and the SFR of the most massive one ($10.75<\log M_* $) at $z=0$ is $\sim 1.6$ dex for star forming galaxies, and slightly higher for quiescent galaxies. 

A similar trend in Figure \ref{fig:ssfr_halomass_qs_mass} is found for the SSFR of the star forming sample, while the trend does not appear clear for the quiescent one. However, it must be noted that the average 
gap in SSFR for star forming galaxies at $z=0$ is $\sim 0.5$ dex. Moreover, less massive galaxies are those more star forming among all, in good agreement with previous studies (e.g., \citealt{peng10}), and 
in general with the downsizing scenario for which less massive galaxies quench on longer timescales (e.g., \citealt{popesso11,sobral11,fossati17,pintoscastro19}; Rhee et al. 2019, in prep). 

We now want to see whether the stellar mass quenching is dependent on the definition of the environment and so, instead of using the halo mass as proxy of the environment, we plot the same 
quantities as a function of clustercentric distance. This is done in Figure \ref{fig:distance_mass}, which shows the SFR (left panel) and SSFR (right panel) of star forming (solid lines) and quiescent 
(dashed lines) galaxies as a function of distance from the halo center, for galaxies in different stellar mass bins, at $z=0$.  With respect to Figure \ref{fig:sfr_halomass_qs_mass}, where the environment 
was defined as the mass of the cluster in which galaxies reside, the general trends and average gaps between the two extreme ranges in stellar mass do not change much (exception made for the SSFR of quiescent 
galaxies). The results found in Figures \ref{fig:sfr_halomass_qs_mass} and \ref{fig:distance_mass} strongly suggest that at fixed environmental conditions, may be them given by the typical halo mass or 
clustercentric distance, the SFR and SSFR depend on stellar mass. The predictions of our model agree well with many observational results in the literature (e.g. \citealt{muzzin12,koyama13,lagana18}). We 
will compare our results with previous findings and fully discuss their implications in Section \ref{sec:discussion}.

\section{Discussion}
\label{sec:discussion}

The main goal of this work is to study the roles of mass and environmental quenching separately. Our analytic model has been developed with the purpose of describing the evolution of the galaxy stellar mass 
function from high to low redshift and, at the same time, to give a reasonable prediction of the evolution of the SFR-$M_*$ relation which agrees with that of the SMF. The model follows the star formation 
history of each galaxy and treat them differently depending on their type (central or satellite) and on their quenching timescale (which is mass and redshift dependent). Hence, the effects of environment 
and mass are robustly considered. In simple words, central galaxies actively form stars for a given time that depends on their quenching timescale, but when they become satellites, they keep forming stars as 
they are active centrals for a few Gyr, and experience a rapid quenching later on. Such a model predicts different roles for mass and environment in quenching galaxies, that are important at different redshifts and 
in a non-linear relation with the galaxy stellar mass. Below we discuss them and their implications according to the results obtained in the analysis done in sections \ref{sec:env_quenching} and \ref{sec:mass-quenching}.

In Figure \ref{fig:sfrmass_qs_halomass} and \ref{fig:ssfrmass_qs_halomass} we have analysed the dependence with time of the SFR-$M_*$ (Fig. \ref{fig:sfrmass_qs_halomass}) and SSFR-$M_*$ (Fig. \ref{fig:ssfrmass_qs_halomass})
relations for galaxies in different environments defined as the halo mass, from $z=1.5$ to $z=0$. Our results are consistent with a scenario where the enviromental processes play a marginal effect in galaxy 
quenching, at any time, or they are very rapid in such a way that the net environmental quenching is not seen. This scenario is supported by a number of observational achievements
(e.g., \citealt{peng10,sobral11,muzzin12,koyama13,lagana18}). 

Very recently \cite{lagana18}, who analyzed the relation between the SFR and SSFR as a function of environment and stellar mass for galaxies in cluster at intermediate redshift ($0.4<z<0.9$), found no dependence of 
the star formation activity on environment. Moreover, they suggest that for cluster galaxies in that redshift range, mass must be the main driver of quenching. \cite{muzzin12} in one of their main conclusions state 
that, in the redshift range they probed ($z\sim 1$), ``the stellar mass is the main responsible for determining the stellar populations of both star forming and quiescent galaxies, and not their environment''. 
In their work they used the clustercentric distance as a proxy of environment, and so, according to their Fig. 10, where they plot the SSFR as a function of the distance from the centre of the cluster (they probed also 
longer distances where galaxies can be classified as being in the field), quenching is not sensitive to the particular location of a given galaxy. 

Their conclusion is supported by other works, such as \cite{wetzel12,darvish16,lagana18}. \cite{darvish16} used a sample of star forming and quiescent galaxies in the COSMOS field at $z<3$, and studied the role of 
environment and stellar mass on galaxy properties, in particular the evolution of the SFR and SSFR with overdensity (as a proxy of the environment) as a function of redshift. For all galaxies, although at $z>1$ the 
SFR and SSFR do not depend on the overdensity (i.e. no environmental dependence), at lower redshift they strongly do. However, once star forming systems are isolated, no clear dependence on the 
overdensity is seen at any redshift. This is in good agreement with our results, and in general with a picture where the environment does not influence the star formation activity of star forming galaxies, but it can 
increase the probability of a given galaxy to become quiescent. Indeed, it has been pointed out by many authors (\citealt{patel09,peng10,darvish16,pintoscastro19}) that the fraction of quiescent galaxies strongly 
depends on the environment, but the SFR and SSFR of star forming galaxies are independent of environment (\citealt{muzzin12,wetzel12,koyama13,darvish16,lagana18}; this work).

It must be noted, however, that there are claims for an environmental dependence on the SFR also for star forming galaxies (e.g., \citealt{vonderlinden10,patel11,woo13,tran15,schaefer17}). As discussed previously, part of 
the tension can be attributed to different reasons. We have already cited the importance of the method of separating star forming from quiescent galaxies (Section \ref{sec:results}), but different SFR indicators, the 
selection of the environment, and cosmic variance might play a non-negligible role. In particular the environment itself , which is still probably one the most undefined (or ill defined) galaxy property in astrophysics. 
Its definition ranges from halo mass in which galaxies reside, to clustercentric distance (or normalised by the virial radius of the cluster) and local overdensity within the Nth-nearest neighboor.
Another possible source for the disagreement between the predictions of our model and the results of the studies quoted above might be found in the sensitivity of the model parameters, especially in the delay time 
and quenching timescale ($\tau_s$) of satellites. On this regard, we ran the model by applying reasonable variations (up to $\pm 30\%$ on the delay time $t_{delay}$, and $\pm 50\%$ on the random fraction $f_{\tau}$), 
finding no appreciable difference with the results obtained in this analysis. However, higher percentages would change the evolution of the predicted SMF and worsen its comparison with the observed one, which would go 
against the main goal of our model.

It appears clear from our analysis that stellar mass is the main driver of galaxy quenching, at any redshift probed in this study. This is the main conclusion of our work, which fits well with the growing observational 
evidence that supports it, at least down to redshift $z\sim 0.5$ (e.g., \citealt{lagana18}). The novelty of this paper is to extend mass quenching as the primary mode of shutting down star formation in star forming galaxies 
down to the present time.

Before concluding, it is important to quote a number of observational results that imply a connection between mass and environmental quenching. These two modes of quenching have been treated as separable by many 
authors (e.g., \citealt{peng10,muzzin12}, this work and many others), but there is a growing consensus (mainly among observers) for which environmental quenching is mass dependent in very dense environments such as 
the cores of galaxy clusters (\citealt{balogh16,darvish16,kawinwanichakij17,papovich18,pintoscastro19}). All these quoted works found a mutual dependence between the mass and environmental quenching efficiencies, from $z>1$ 
(\citealt{kawinwanichakij17}), to $z\sim 0.4$ (\citealt{pintoscastro19}). The analysis done in this work does not allow us to either confirm or prove wrong such a (important) statement. In principle, if mass and 
environmental quenching are mutually dependent, this should be seen in Figure \ref{fig:distance_mass}, where the SFR/SSFR of star forming galaxies in each stellar mass bin should depend on the distance from the cluster 
core, and they do not. However, mass and environmental quenching efficiencies have well precise definitions. The environmental quenching efficiency is usually defined as the increase of the fraction of quiescent galaxies 
at a given distance from the cluster centre with respect to the field, normalised by the fraction of star forming galaxies in the field. The mass quenching efficiency is defined in a similar way by means of a characteristic 
mass at which almost all galaxies at a given distance bin are star forming. The information in Figure \ref{fig:distance_mass} is then not enough to make a fair comparison with the works cited above. We aim to address this point 
with a full analysis in a forthcoming paper.

\section{Conclusions}
\label{sec:conclusions}
We have studied the roles of stellar mass and environment in quenching galaxies by taking advantage of an analytic model of galaxy formation. The model was set in order to match the evolution of 
the global stellar mass function from high to low redshift and, at the same time, to give reasonable predictions of the star formation history of galaxies. From the analysis done in this work we 
can conclude the following:
\begin{itemize}
 \item The SFR/SSFR-$M_*$ relations are independent of the environment at any redshift probed, $0<z<1.5$, for both star forming and quiescent galaxies.
 \item The SFR-$M_{halo}$ relation strongly depends on stellar mass at any redshift probed, for both star forming and quiescent galaxies.
 \item The SSFR-$M_{halo}$ relation strongly depends on stellar mass at any redshift probed for star forming galaxies, while the trend is not clear for the quiescent sample.
 \item Overall, less massive galaxies are more star forming, in agreement with the downsizing scenario for which less massive galaxies quench on longer timescales.
 \item The SFR and SSFR are strongly dependent on stellar mass even when the distance from the cluster core is used as a proxy for the environment (rather than the halo mass).
\end{itemize}
All these conclusions put together draw a picture where stellar mass is the main driver of galaxy quenching at any redshift, not only at $z>1$ as generally claimed in the literature. The role of 
environment is marginal: environmental processes must act very fast such that they do not have an effect on the star formation activity of star forming galaxies, but can increase the probability of a 
galaxy to become quiescent. 

In a forthcoming paper we will address the point of the mutual dependence of the mass and environment quenching efficiencies by looking directly at the star forming and quiescent fractions in galaxy 
clusters, and compare the predictions of our model with the newest observational evidence.

\section*{Acknowledgements}
This work is supported by the National Key Research and Development Program of China (No. 2017YFA0402703), 
by the National Natural Science Foundation of China (Key Project No. 11733002), the Korean National 
Research Foundation (NRF-2017R1A2A05001116), and the NSFC grant (11825303, 11861131006). E.C. acknowledges 
support from the Faculty of the European Space Astronomy Centre (ESAC) - Funding reference 497.

\label{lastpage}

\end{document}